\documentclass[twocolumn,showpacs,preprintnumbers,amsmath,amssymb,superscriptaddress]{revtex4-2}

\usepackage{epsf}
\usepackage{graphicx}
\usepackage{sidecap}
 \usepackage{soul}
 \usepackage{array}
 \usepackage{amsmath}
 \usepackage{amssymb}
 \usepackage{dcolumn}
 \usepackage{epstopdf}
 \usepackage{bm}
 \usepackage{amsmath}
 
\usepackage{gensymb}
\usepackage{color}
\usepackage{hyperref}
\usepackage{soul}
\sethlcolor{green}
\usepackage{bbding}
\hypersetup{
    colorlinks=true,
    citecolor=red,
    linkcolor=red,
    filecolor=blue,   
    urlcolor=blue,
}
 
\def \beq {\begin{equation}}
\def \eeq {\end{equation}}
\renewcommand{\figurename}{\textbf{Fig.}}
\renewcommand{\thefigure}{{\textbf{\arabic{figure}}}}

\def\bibsection{\refname}
\renewcommand{\refname}{\noindent\textbf{References}}
\begin{document}

\title{Diverse electronic landscape of the kagome metal YbTi$_3$Bi$_4$}
\author{Anup Pradhan Sakhya} \thanks{These authors contributed equally.} \affiliation {Department of Physics, University of Central Florida, Orlando, Florida 32816, USA} 
\author{Brenden R. Ortiz} \thanks{These authors contributed equally.} \affiliation {Materials Science and Technology Division, Oak Ridge National Laboratory, Oak Ridge, Tennessee 37830, USA}
\author{Barun Ghosh} \thanks{These authors contributed equally.} \affiliation{Department of Physics, Northeastern University, Boston, Massachusetts 02115, USA}
\affiliation{Quantum Materials and Sensing Institute, Northeastern University, Burlington, Massachusetts 01803, USA}
\author{Milo Sprague} \affiliation{Department of Physics, University of Central Florida, Orlando, Florida 32816, USA} 
\author{Mazharul Islam Mondal} \affiliation{Department of Physics, University of Central Florida, Orlando, Florida 32816, USA} 
\author{Matthew Matzelle} \affiliation{Department of Physics, Northeastern University, Boston, Massachusetts 02115, USA}
\affiliation{Quantum Materials and Sensing Institute, Northeastern University, Burlington, Massachusetts 01803, USA}
\author{Iftakhar Bin Elius} \affiliation{Department of Physics, University of Central Florida, Orlando, Florida 32816, USA} 
\author{Nathan Valadez} \affiliation{Department of Physics, University of Central Florida, Orlando, Florida 32816, USA} 
\author{David G. Mandrus} \affiliation{Department of Materials Science and Engineering, The University of Tennessee, Knoxville, Tennessee 37996, USA}
\affiliation {Department of Physics and Astronomy, University of Tennessee Knoxville, Knoxville, Tennessee 37996, USA}
\affiliation{Materials Science and Technology Division, Oak Ridge National Laboratory, Oak Ridge, Tennessee 37831, USA}
\author{Arun Bansil} \affiliation{Department of Physics, Northeastern University, Boston, Massachusetts 02115, USA}
\affiliation{Quantum Materials and Sensing Institute, Northeastern University, Burlington, Massachusetts 01803, USA}
\author{Madhab Neupane} \thanks{Corresponding author:\href{mailto:madhab.neupane@ucf.edu}{madhab.neupane@ucf.edu}}\affiliation{Department of Physics, University of Central Florida, Orlando, Florida 32816, USA}

\date{\today}

\begin{abstract}
{\indent Kagome lattices have emerged as an ideal platform for exploring exotic quantum phenomena in materials. Here, we report the discovery of Ti-based kagome metal YbTi$_3$Bi$_4$ which we characterize using angle-resolved photoemission spectroscopy (ARPES) and magneto-transport, in combination with density functional theory calculations. Our ARPES results reveal the complex fermiology of YbTi$_3$Bi$_4$ and provide spectroscopic evidence of four flat bands. Our measurements also show the presence of multiple van Hove singularities originating from Ti 3\textit{d} orbitals and a linearly-dispersing gapped Dirac-like bulk state at the $\overline{\text{K}}$ point in accord with our theoretical calculations. Our study establishes YbTi$_3$Bi$_4$ as a platform for exploring exotic phases in the wider $Ln$Ti$_3$Bi$_4$ ($Ln$= lanthanide) family of materials.} 

\end{abstract}

\maketitle  
\begin{center}\textbf{I. INTRODUCTION}
\end{center}

The kagome motifs, which are composed of two-dimensional honeycomb networks of alternating corner-sharing triangles, are attracting enormous interest as they provide a fascinating platform for exploring the emergence of many topological and correlated electronic phenomenon \cite{Syozi, Zhou, Neupert}. The Kagome structure allows the possibility of flat bands, Dirac-like dispersions, and van Hove singularities (vHSs), resulting in a complex interplay of geometry, topology, and electronic correlation effects \cite{Guo, Tang, Han, Ye, Lin, Yin1, Yin2, Kang1, Kang2, Nirmal, Li, Regmi, Yinkagome, Balents, Yang}. For example, a class of kagome superconductors, \textit{A}V$_3$Sb$_5$ (\textit{A} = K, Rb, and Cs),  was discovered recently and found to host nontrivial band topology, charge density wave order and a superconducting ground state \cite{Ortiz1, Ortiz2, Ortiz3, Ortiz4, Zhao, Hu, Jiang, Hao, LeiRb135, LuoK135, Wilson}.

\indent Intertwining of magnetism with the electronic states in the kagome materials has yielded notable discoveries such as the massive Dirac fermions in Fe$_3$Sn$_2$ \cite{Ye}, antiferromagnets Mn$_3$Sn and Mn$_3$Ge that showcase giant anomalous Hall effect \cite{Nakatsuji, Nayak} and the Weyl semimetal state in Co$_3$Sn$_2$S$_2$ \cite{Liu, Chenweyl}.  The \textit{R}Mn$_6$Sn$_6$ kagome family (\textit{R}= rare earth element) has garnered special attention due to its unique magnetic tunability and large Berry curvature fields \cite{Yin2, Li, Ma, Wang, Gu, Asaba, Zeng, Y16, Dhakal, Wangymsn6, Kabir, Lv}. TbMn$_6$Sn$_6$ hosts massive Dirac fermions at the K point owing to Landau quantization and the presence of Landau fan diagram \cite{Yin2, Ma}. Flat bands and Dirac fermions have been reported in RMn$_6$Sn$_6$ (R =Gd-Dy, Y) \cite{Wang, Gu, Li} family, topological Hall effect has been observed in YMn$_6$Sn$_6$, ErMn$_6$Sn$_6$, and HoMn$_6$Sn$_6$ \cite{Y16, Dhakal, Wangymsn6, Kabir}, and a large anomalous Hall effect has been reported in LiMn$_6$Sn$_6$ \cite{Felser}. 

\indent Recently, two kagome metals, YbV$_3$Sb$_4$ and EuV$_3$Sb$_4$ were reported, introducing a new family of vanadium-based kagome materials \cite{Ortiz5}. It was found that YbV$_3$Sb$_4$ is a Pauli paramagnet with no clear thermodynamic phase transition between 300 K and 60 mK, whereas EuV$_3$Sb$_4$ displays a ferromagnetic transition below T$_c$ of roughly 32 K and a signature of spin-texture under low-fields. The larger \textit{Ln}M$_3$X$_4$ (MX: VSb, TiBi) family is growing rapidly \cite{alexander, BrendenLn134}, providing a new kagome materials platform.

\begin{figure*} 
	\includegraphics[width=17.8cm]{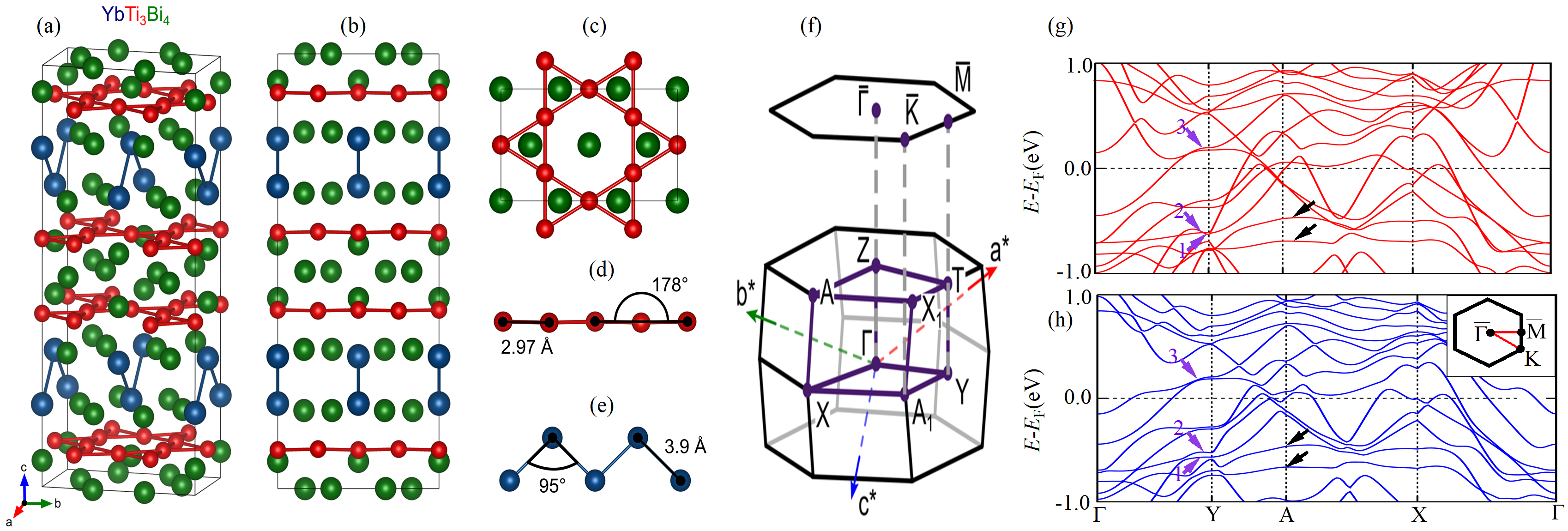} 
   \vspace{-1ex}
	\caption{Crystal structure and DFT results for YbTi$_3$Bi$_4$. (a)  Crystal structure of YbTi$_3$Bi$_4$ where the blue, red, and green solid spheres denote Yb, Ti, and Bi atoms, respectively. The crystal structure is orthorhombic \textit{(Fmmm)} with a zigzag sublattice of Yb atoms. (b) The kagome lattice, which is formed by the Ti atoms, (c) is slightly distorted and it is not coplanar. (d) A view of the structure perpendicular to the \textit{b-c} plane, which highlights the relatively small out-of-plane distortion of about ($\sim$ 2\degree). (e) Zig-zag chain formed by the Yb atoms. (f) Bulk Brillouin zone (BZ) and its projection onto the (001) surface BZ; high symmetry points are marked. (g) DFT based energy bands along the various high-symmetry directions without the inclusion of spin-orbit coupling (SOC). (h) Same as panel (g) except that these results include SOC. Purple and black arrows in (g) and (h) point to vHSs and flat bands, respectively.}
\label{fig1}
\end{figure*}

Here we discuss our comprehensive investigation of the electronic structure of the YbTi$_3$Bi$_4$ kagome material. Our approach involves magneto-transport and angle-resolved photoemission spectroscopy (ARPES) measurements, in combination with parallel density-functional theory (DFT) based calculations. YbTi$_3$Bi$_4$, like its structural analog YbV$_3$Sb$_4$ is a nonmagnetic (Yb$^{2+}$) kagome metal, which does not exhibit any electronic or structural phase transitions down to 60 mK \cite{BrendenLn134}. Our ARPES measurements on YbTi$_3$Bi$_4$ indicate the presence of several flat bands spanning a large area of the Brillouin zone (BZ), multiple van Hove singularities at the $\overline{\text{M}}$ point, and a linear Dirac-like state at the $\overline{\text{K}}$ symmetry point, and are consistent with our DFT calculations. Our study establishes YbTi$_3$Bi$_4$ as a fertile playground for exploring exotic physics in an easily exfoliable Ti-based nonmagnetic kagome material.\\ 


\begin{figure*} 
	\includegraphics[width=18cm]{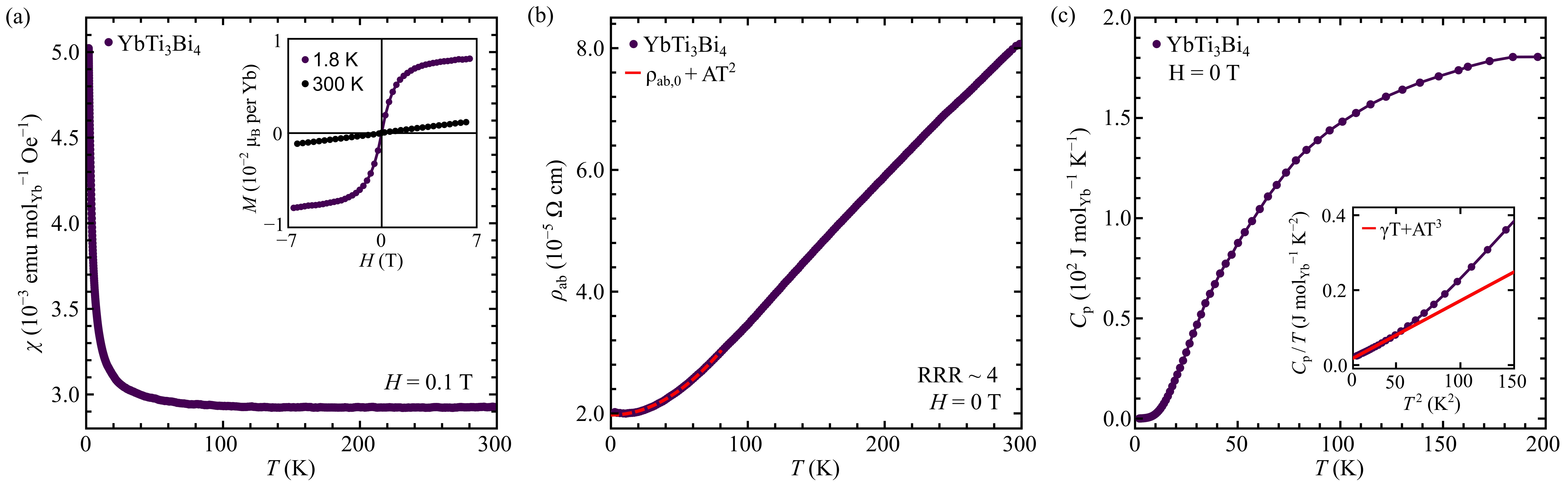} 
    \vspace{-1ex}
	\caption{Bulk electronic properties and characterization of a YbTi$_3$Bi$_4$ single crystal specimen. (a) Magnetic susceptibility ($\chi$) vs. temperature under a magnetic field of H = 0.1 T. Inset shows magnetization as a function of magnetic field M(H) up to 7 T at 300 K and 1.8 K. (b) Temperature-dependent zero-field electrical resistivity. (c) Temperature dependence of the specific heat (\textit{C$_p$}) over 2-300 K. Inset highlights \textit{C$_p$}/T versus \textit{T}$^2$ in the low temperature regime.}
\label{fig2}
\end{figure*}
 
\begin{center}\textbf{II. EXPERIMENTAL AND COMPUTATIONAL DETAILS}
\end{center}
\indent YbTi$_3$Bi$_4$ single crystals were grown through bismuth self-flux. Elemental reagents of Yb (Alfa 99.9\%), Ti (Alfa 99.9\% powder), and Bi (Alfa 99.999\% low-oxide shot) were combined in 2:3:12 ratio into 2~mL Canfield crucibles fitted with a catch crucible and a porous frit \cite{Canfield}. The crucibles were sealed under approximately 0.7~atm of argon gas in fused silica ampoules. Each composition was heated to 1050 \degree C at a rate of 200 \degree C/hr. Samples were allowed to thermalize and homogenize at 1050 \degree C for 12-18~hr before cooling to 500 \degree C at a rate of 2\ degree C/hr. Excess bismuth was removed through centrifugation at 500 \degree C.\\ 

Synchrotron-based ARPES experiments were performed at the ALS beamlines 10.0.1.1 and 4.0.3 at a temperature of 15 K equipped with R4000 and R8000 hemispherical electron analyzers. Additional ARPES experiments were carried out at the Stanford Synchrotron Radiation Lightsource end station 5-2 at 15 K. The angular and energy resolution were set at better than 0.2\degree and 15 meV, respectively. High-quality single crystals were cut into small pieces and mounted on a copper post using silver epoxy. Ceramic posts were attached on top of the samples. To avoid oxidation,  samples were prepared in a glove box, and loaded into the main chamber which was cooled and pumped down for a few hours. Measurements were carried out at 11 K. Pressure in the UHV chamber was maintained at around 10$^{-11}$ Torr.\\ 

Electronic structure calculations were performed within the density functional theory (DFT) framework using a plane wave basis set as implemented in the Vienna Ab initio Simulation Package (VASP) \cite{Kohn,kresse1996efficient}. Standard projector-augmented-wave (PAW) pseudopotentials were used \cite{kresse1999ultrasoft}. Kinetic energy cutoff for the plane-wave basis was set to 500 eV. A $\Gamma$-centered 10 $\times$ 10 $\times$ 10 \textit{k} mesh was used for \textit{k}-space integrations. The SCAN (strongly-constrained and appropriately-normed) functional was used to treat exchange-correlation effects \cite{SunPRL, SunNature}. Yb\_3 pseudopotential was used, where the Yb 4\textit{f} electrons are taken to be core electrons. The lattice parameters were fully relaxed by optimizing atomic positions until the force on each atom became less than 0.001 eV\AA. The relaxed lattice parameters are a=5.862 \AA, b=10.385 \AA, c=24.838 \AA, which are very close to the corresponding experimental values.\\ 

\maketitle  
\begin{center}\textbf{III. RESULTS AND DISCUSSION}
\end{center}
YbTi$_3$Bi$_4$ crystallizes in the orthorhombic space group F\textit{mmm} (No. 69) with a = 5.91(4) \AA, b = 10.3(4) \AA, c =24.9(4) \AA, and $\alpha$ = $\beta$ = $\gamma$ = 90\degree. Figure (1a-b) shows that the structure consists of Ti kagome layers stacked along the \textit{c}-axis between the Bi and Yb layers. To emphasize the kagome lattice formed by the Ti atoms and the zig-zag chains formed by the Yb atoms, we highlight only the Ti-Ti, and Yb-Yb bonds in Figure 1. Kagome layers formed by the Ti atoms are better visualized in Figure 1(c), where each unit cell consists of four kagome layers. However, unlike \textit{A}V$_3$Sb$_5$, the kagome layers formed by the Ti atoms in YbTi$_3$Bi$_4$ exhibit slight buckling, see Figure 1(b,d). Figure 1(e) illustrates the zig-zag chains formed by the Yb atoms extending in the \textit{a}-direction. The Yb-Yb distance in the chains is approximately 3.9 \AA, which is significantly shorter than the nearest neighbor Yb-Yb interchain distance of approximately 5.7 \AA. 

\indent The bulk Brillouin zone (BZ) and its projection on the (001) surface (labeled with the pseudo-hexagonal $\overline{\text{M}}$, $\overline{\text{K}}$, and $\overline{\Gamma}$ nomenclature) is presented in Figure 1(f). To gain insight into the electronic structure of YbTi$_3$Bi$_4$, we performed band structure calculations without [Figure 1(g)] and with the inclusion of the spin-orbit coupling effects [Figure 1(h)]; see Supplementary Note 1 and Supplementary Figure 1 for orbital-resolved band structures. Multiple bands are seen to cross the Fermi level (E$_{\text{F}}$) consistent with the metallic nature of YbTi$_3$Bi$_4$. Effects of SOC are generally small, except for the Dirac-like crossing along the $\Gamma$-X direction which becomes gapped with the inclusion of SOC. We observe three vHS points at Y (purple arrows) along with the presence of dispersionless bands below the E$_{\text{F}}$ (black arrows), indicating a high degree of localization of the associated electronic states. The dispersionless or ``flat" bands are seen to extend over a large portion of the BZ. The flat bands at $\sim$ - 0.5 eV and $\sim$ - 0.7 eV predominantly arise from the kagome-based Ti \textit{d$_{xy}$} and \textit{$d_{x^{2}-y^{2}}$} orbitals, see Supplementary Note 1 and Supplementary Figure 1.

\indent For experimental work, we grew high-quality single crystals of YbTi$_3$Bi$_4$ using the flux method. Temperature-dependent molar magnetic susceptibility $\chi(T)$ =$\textbf{M/H}$ measured under magnetic field of \textbf{H} = 0.1 T is presented in Figure 2(a) for \textbf{H} $\parallel$ \textit{c}. Essentially temperature-independent Pauli paramagnetism is seen with a weak Curie tail from the impurity spins. Extremely weak nature of the susceptibility (10$^{-3}$ emu Oe$^{-1}$ mol$^{-1}$) is evident from Figure 2(a), which does not display any signature of a bulk magnetic moment. No qualitative difference is observed when \textbf{H} is applied $\parallel$ versus $\perp$ to the \textit{c}-axis. The inset compares magnetization as a function of the magnetic field at 300 K and 1.8 K, and shows no saturation up to 7 T. Note that the scale of the \textbf{M} versus \textbf{H} plot is of the order of 10$^{-2}$ $\mu_B$ per Yb, consistent with the polarization of impurity spins. 

\indent Figure 2(b) shows the electrical resistivity $\rho$ as a function of temperature, where the current is flowing in the \textit{ab}-plane. The typical metallic behavior is seen until the lowest temperature of 2 K used in the measurements. The residual resistivity is approximately 20 $\mu\Omega$ cm and the residual resistivity ratio is around 4. The resistivity data at low temperatures ($<$ 80 K) is modeled using the Fermi liquid picture, which yields the simple quadratic form, $\rho$ = $\rho_0$+$aT^2$ (red dotted line), where $\rho_0$  = 19.7 $\mu\ohm$ cm and \textit{a} = 1.6 $\times$ 10$^{-3}$ $\mu\ohm$ cm K$^{-2}$. We also measured the specific heat (C$_p$) of the sample from 200 K down to 2 K without magnetic field [Figure 2(c)]. We have fitted the C$_p$ data over the limited temperature range of 2-12 K using the Sommerfeld model. 
\indent Recall that the C$_p$ for nonmagnetic metals consists of of electronic C$_e$ and lattice C$_{ph}$ contributions, and at low temperature it is given by: C$_p$(T)=$\gamma$\textit{T}+$\beta$\textit{T}$^{3}$, where C$_e$ = $\gamma$\textit{T} and C$_{ph}$ = $\beta$\textit{T}$^{3}$. Here, $\gamma$ is the Sommerfeld coefficient, and $\beta$ = \text{12$\pi^{4}$$NR/5$$\Theta_D^3$} with $N$, $R$, and $\Theta_D$ representing the number of atoms per unit cell, the ideal gas constant, and the Debye temperature, respectively. The inset of Figure 2(c) shows that the C$_p$/(T) scales linearly with \textit{T}$^{2}$ at low temperatures. The resulting least-squares fit yields the parameters $\gamma$ = 1.78 $\times$ 10$^{-2}$ J mol$^{-1}$ K$^{-2}$ and $\beta$ = 1.46 $\times$ 10$^{-3}$ J mol$^{-1}$ K$^{-4}$. These results indicate that YbTi$_3$Bi$_4$ is a nonmagnetic kagome metal with no clear thermodynamic phase transitions up to 300 K. 

\begin{figure*} 
	\includegraphics[width=17cm]{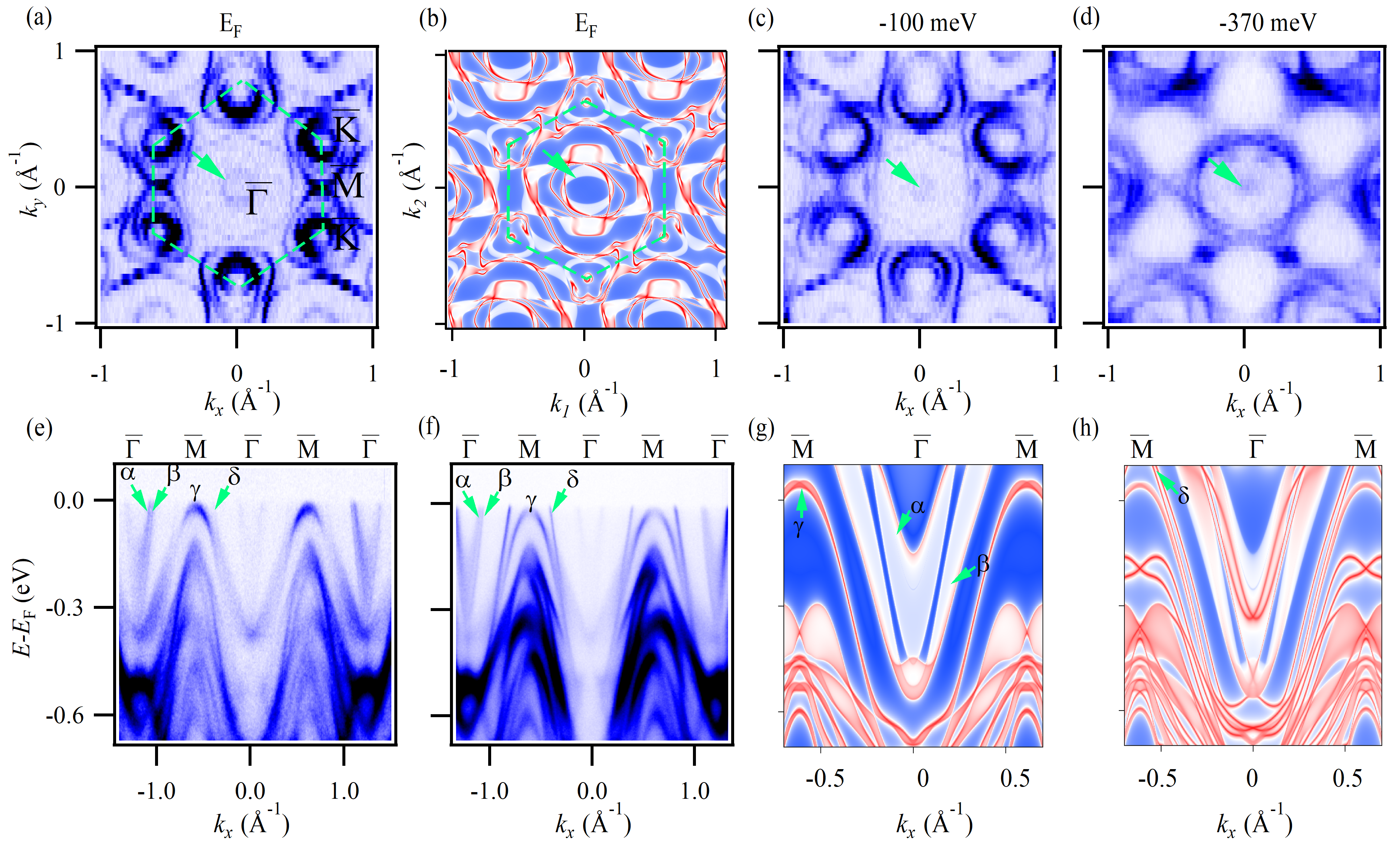} 
    \vspace{-1ex}
	\caption{Fermi surface (FS) and band-dispersion along the $\overline{\text{M}}$--$\overline{\Gamma}$--$\overline{\text{M}}$ high-symmetry line. (a) ARPES measured FS using a photon energy of 90 eV. High-symmetry points are labeled. (b) DFT calculated FS. (c-d) Constant-energy contours at various binding energies as noted on the top of each plot. (e) Experimental band dispersion measured along the $\overline{\Gamma}$--$\overline{\text{M}}$--$\overline{\Gamma}$--$\overline{\text{M}}$--$\overline{\Gamma}$ high-symmetry line with a photon energy of 90 eV. (f) Same as (e) except that the measurements are with a photon energy of 75 eV. (g) Projected bulk band structure along the $\overline{\text{M}}$--$\overline{\Gamma}$--$\overline{\text{M}}$ high-symmetry line. (h) same as (g) except that this panel includes the surface states.}
\label{fig3}
\end{figure*}

\indent We turn now to discuss our ARPES measurements from YbTi$_3$Bi$_4$. Note that our DFT calculations indicate that the density of states (DOS) at the E$_{\text{F}}$ in YbTi$_3$Bi$_4$ is dominated by the Ti and Bi orbitals. With the Yb-orbitals lying away from the E$_{\text{F}}$, ARPES is expected to show an apparent 6-fold symmetry, although the underlying structure is orthorhombic. We focus first on the Fermi surface (FS) topology. Figure 3(a) presents the FS measured using a photon energy of 90 eV as a function of \textit{k$_x$} and \textit{k$_y$}. Several pockets can be seen at the E$_{\text{F}}$, with a circular pocket near the BZ center $\overline{\Gamma}$, a pocket with hexagonal symmetry which is anticipated due to the underlying symmetry of the kagome lattice along the $\overline{\Gamma}$--$\overline{\text{M}}$ direction, and two pockets around the $\overline{\text{K}}$ point, where the inner pocket seems to show a circular shape and the outer pocket is triangular; see Supplementary Note 2 and Supplementary Figure 2(a) for additional results of FS measurements. The DFT based bulk FS including surface states, presented in Figure 3(b), is in good agreement with our experimental results, see Supplementary Note 2 and Supplementary Figure 2(b) for bulk FS without the surface states. Discrepancies between the theory and experiment may be attributed to effects of the ARPES matrix element \cite{AB1, AB2} and the intrinsic limitations of the DFT in capturing electronic correlations \cite{Sakhyasmbi, anupndsb, anupweyl}. 

\indent ARPES measured constant energy contours (CECs) are presented in Figure 3(c) and Figure 3(d) at various binding energies. With increasing binding energy, the circular pocket at $\overline{\Gamma}$ contracts to a point-like feature around $\sim$ -370 meV indicating its electron-like nature, whereas the pocket at $\overline{\text{K}}$/$\overline{\text{M}}$ increases in size with increasing binding energy, indicating its hole-like nature. Note that the underlying orthorhombic structure naturally imparts small distortions to the kagome lattice, which result in the apparent 6-fold symmetry in the ARPES FS and CECs. Although we refer to the data using the hexagonal nomenclature, the underlying symmetry is 2-fold. 

\indent Band dispersions along the $\overline{\Gamma}$--$\overline{\text{M}}$--$\overline{\Gamma}$--$\overline{\text{M}}$--$\overline{\Gamma}$ high-symmetry line using photon energies of 90 eV and 75 eV are shown in Figure 3(e) and Figure 3(f), respectively. In agreement with our measured FS and CECs, an electron-like pocket can be seen at the $\overline{\Gamma}$ point. However, a closer examination of the band dispersion at $\overline{\Gamma}$ in the second BZ reveals the presence of two distinct pockets. This is also evident in the band-dispersion plot obtained with 75 eV photons along the $\overline{\Gamma}$--$\overline{\text{M}}$--$\overline{\Gamma}$--$\overline{\text{M}}$--$\overline{\Gamma}$ line, see, Figure 3(f). [For detailed analysis of the bands, labeled $\alpha$ and $\beta$, see Supplementary Note 3 and Supplementary Figure 3(a-b)]. We also obtained momentum distribution curves (MDCs) by integrating over an energy window of 10 meV below the Fermi level around the $\overline{\Gamma}$ point in the second BZ and fitted the spectra using two Gaussian peaks (see Supplementary Note 3 and Supplementary Figure 3(c) to further confirm the presence of two peaks ($\alpha$ and $\beta$) at $\overline{\Gamma}$. Additionally, two hole-like bands ($\gamma$ and $\delta$) are visible in Figure 3(e) and Figure 3(f), which exhibit hole-like characteristics consistent with the FSs and CECs presented in Figures 3(a,c-d). Our projected (bulk) DFT calculations in Figure 3(g) are in good agreement with the results of ARPES measurements in Figure 3(e) and Figure 3(f), reproducing the two electron-like pockets $\alpha$ and $\beta$, and the hole-like pocket around the $\gamma$ point, indicating their bulk origin. However, the band denoted as $\delta$ in the ARPES spectra is not reproduced in our DFT calculations. When surface states are included in the calculations, Figure 3(h) shows the appearance of the hole-like pocket $\delta$ (green arrow), suggesting its surface origin.

\begin{figure*} 
	\includegraphics[width=17.7cm]{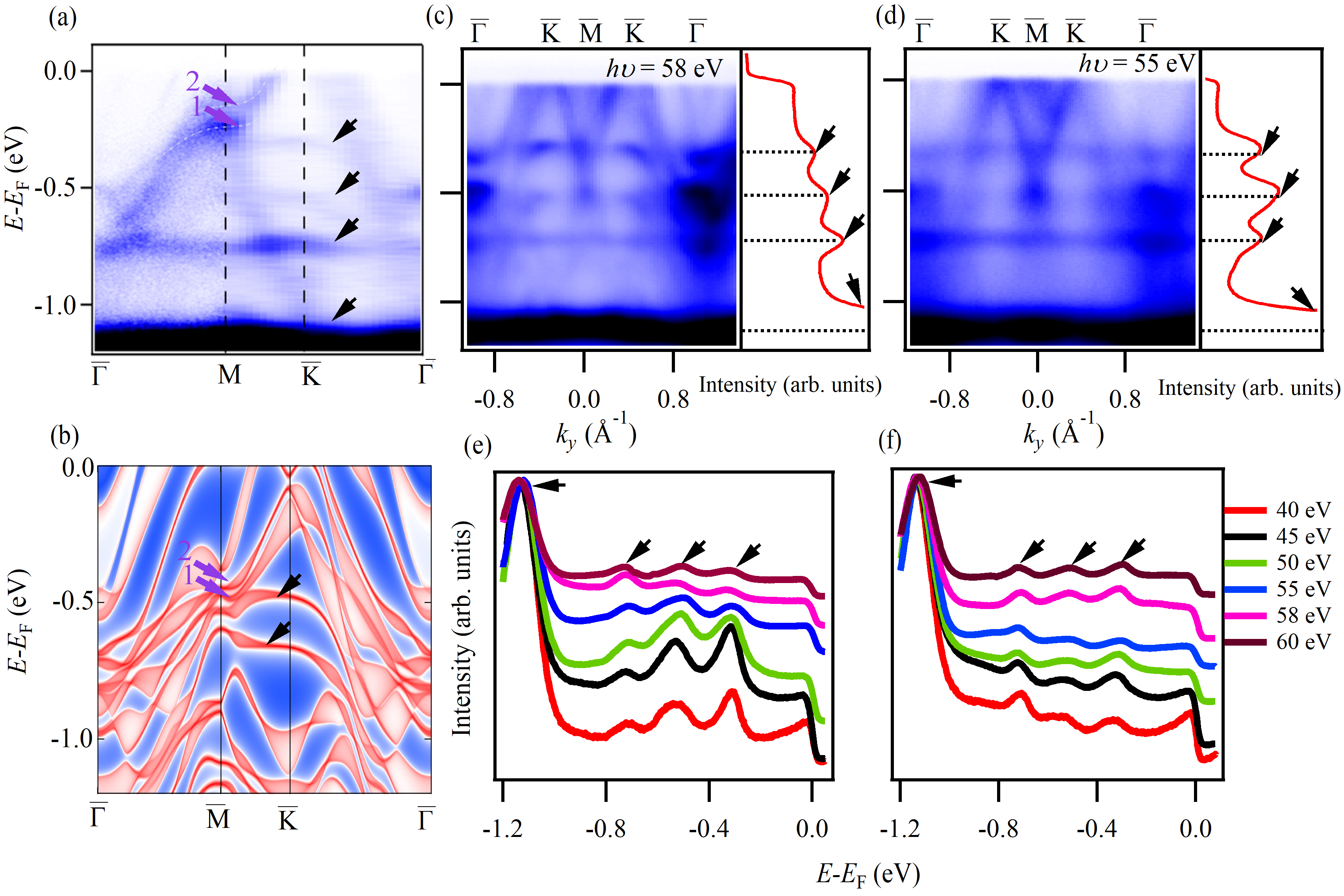} 
    \vspace{-1ex}
	\caption{Visualization of multiple vHSs and flat bands. (a) Experimental photoelectron intensity plot of the band structure along the $\overline{\Gamma}$--$\overline{\text{M}}$--$\overline{\text{K}}$--$\overline{\Gamma}$ line with 90 eV photons. The white-dashed curves serve as guides to the eye to help visualize the two vHSs. Purple and black arrows in panels (a) and (b) point to the vHSs and flat bands, respectively. (b) DFT based bulk band structure along the $\overline{\Gamma}$--$\overline{\text{M}}$--$\overline{\text{K}}$--$\overline{\Gamma}$ line. (c) Experimental band dispersion and the associated integrated energy distribution curves (EDCs) along the $\overline{\Gamma}$--$\overline{\text{K}}$--$\overline{\text{M}}$--$\overline{\text{K}}$--$\overline{\Gamma}$ line measured using 58 eV photons. Peak positions in the EDCs (black arrows) denote flat bands at $\sim$ - 0.3 eV, $\sim$ - 0.5 eV, and $\sim$ - 0.7 eV, with the most intense peak lying at around - $\sim$ 1.1 eV. (d) Same as (c) except these results are for measurements using 55 eV photons. (e) Integrated intensity of the EDCs (over the momentum range of -0.8 \AA${^{-1}}$ to 0.8 \AA${^{-1}}$) along the $\overline{\Gamma}$--$\overline{\text{K}}$--$\overline{\text{M}}$--$\overline{\text{K}}$--$\overline{\Gamma}$ line. The EDCs used to obtain integrated intensities plotted were taken at photon energies ranging from 40 eV to 60 eV as indicated by the color bars on the right side of panel (f). Peak positions in the EDCs (black arrows) correspond to flat bands. (f) Same as (e) except that here the integration involves EDCs along the $\overline{\text{K}}$--$\overline{\Gamma}$--$\overline{\text{K}}$ line.}
\label{fig4}
\end{figure*}

\begin{figure*} 
	\includegraphics[width=12cm]{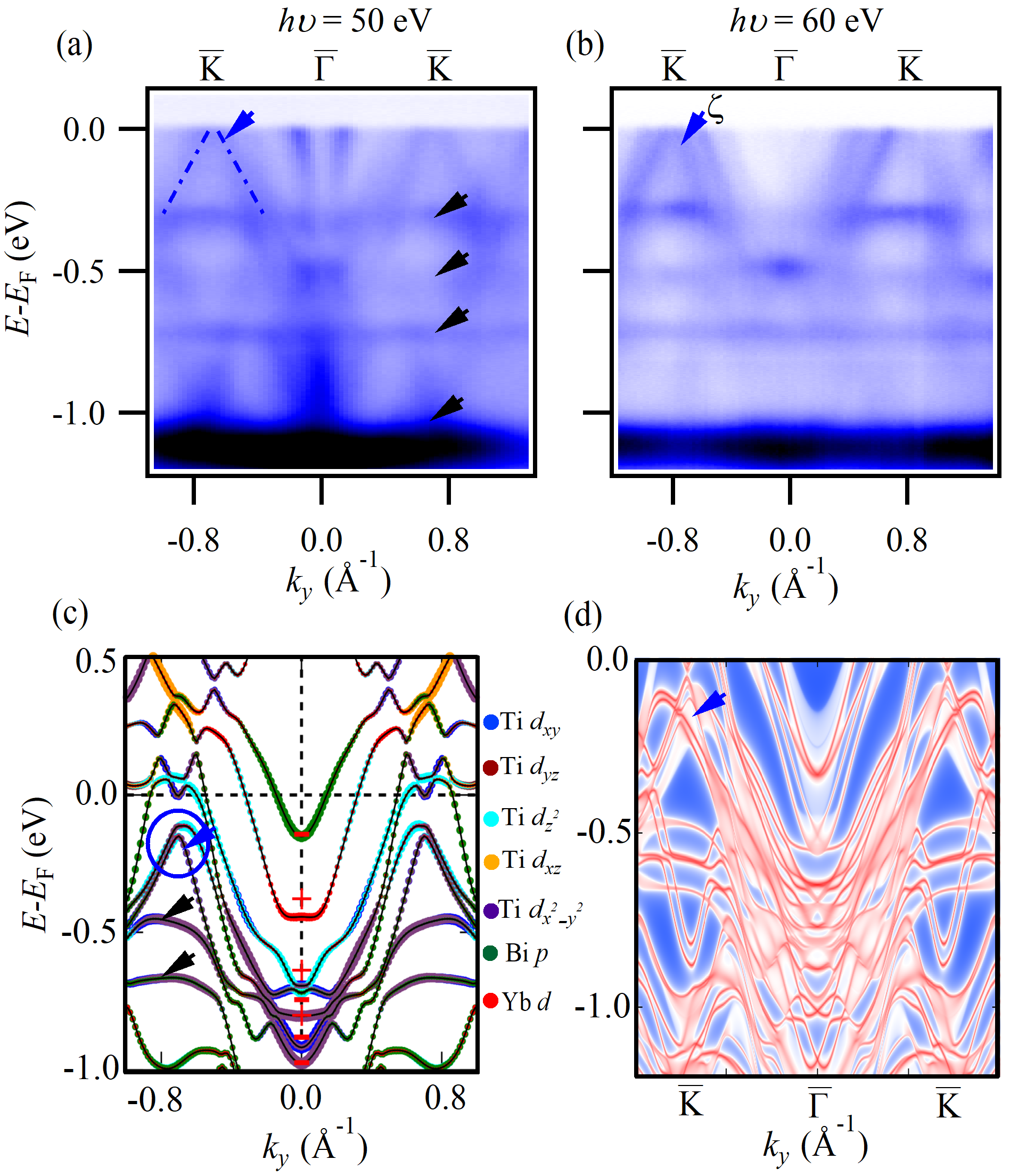} 
    \vspace{-1ex}
	\caption{Observation of a Dirac-like state at the $\overline{\text{K}}$ point. Experimental band dispersions along $\overline{\text{K}}$--$\overline{\Gamma}$--$\overline{\text{K}}$ using a photon energy of (a) 50 eV, and (b) 60 eV. (c) Orbital-projected bulk band structure along $\overline{\text{K}}$--$\overline{\Gamma}$--$\overline{\text{K}}$. Color scale provides relative contributions from various orbitals. (d) DFT based (bulk) projected band structure including the surface states.}
\label{fig5}
\end{figure*} 

\indent Figure 4(a), presents the experimental ARPES band structure along $\overline{\Gamma}$--$\overline{\text{M}}$--$\overline{\text{K}}$--$\overline{\Gamma}$. The most prominent feature is the presence of the vHSs, denoted vHS1 and vHS2 (purple arrows), along with four dispersionless flat bands (black arrows) that extend across a large region of the BZ, which lie at binding energies of $\sim$ - 0.3 eV, $\sim$ - 0.5 eV, $\sim$ - 0.7 eV, and $\sim$ - 1.1 eV. The flat bands at $\sim$ - 0.3 eV and $\sim$ - 0.5 eV are attributed to the destructive interference of wavefunctions within the Ti kagome motif, with interlayer coupling between the two kagome layers resulting in bonding (with a decreased energy at around $\sim$ - 0.5 eV) and antibonding (with an increased energy at around $\sim$ - 0.3 eV) splittings \cite{YCr6Ge6, Ye}. Orbital-resolved DFT calculations suggest that these flat bands predominantly consist of a mixture of Ti \textit{d$_{xy}$} and Ti \textit{$d_{x^{2}-y^{2}}$} orbitals, see Supplementary Note 1 and Supplementary Figure 1. The discrepancy of approximately 200 meV between the DFT calculations and the ARPES results may be due to the neglect of Yb 4\textit{f} states in our calculations \cite{AB1, AB2,Sakhyasmbi, anupndsb, anupweyl}. The flat band observed at approximately $\sim$ - 0.7 eV is associated with Yb surface atoms, while the flat band around  $\sim$ - 1.1 eV originates from bulk Yb$^{2+}$  4\textit{f}$_{5/2}$ states \cite{ZXShen}, see Supplementary Note 4 and Supplementary Figure 4 for additional photon-energy dependent measurements. vHS2 mostly involves Ti \textit{$d_{x^{2}-y^{2}}$} and Ti \textit{$d_{z^{2}}$} orbitals, whereas vHS1 mainly involves the Ti \textit{d$_{xy}$} orbital. The vHS above the E$_{\text{F}}$ seen in DFT calculations, which is not accessible to ARPES, mostly consists of Ti \textit{d$_{xy}$} orbital, see Supplementary Note 1 and Supplementary Figure 1. Except for the flat bands contributed by Yb 4\textit{f} states, the DFT calculated bulk band structure along $\overline{\Gamma}$--$\overline{\text{M}}$--$\overline{\text{K}}$--$\overline{\Gamma}$ in Figure 4(b) reproduces the ARPES measured vHSs (vHS1 and vHS2) and flat bands. The vHSs near the E$_{\text{F}}$ have been implicated in driving exotic physics (e.g. charge ordering and superconductivity) in \textit{A}V$_3$Sb$_5$ (\textit{A}= K, Rb, and Cs) kagome metals. Since the vHSs in YbTi$_3$Bi$_4$ lie away from the E$_{\text{F}}$, this could explain the absence of charge ordering and superconductivity in this material. 

\indent To further examine the flat bands, we present the band dispersion along $\overline{\Gamma}$--$\overline{\text{K}}$--$\overline{\text{M}}$--$\overline{\text{K}}$--$\overline{\Gamma}$ high-symmetry using 58 eV and 55 eV photons in Figure 4(c) and Figure 4(d), respectively. Four dispersionless bands at binding energies of $\sim$ - 0.3 eV, $\sim$ - 0.5 eV, $\sim$ - 0.7 eV, and $\sim$ - 1.1 eV can be seen as before at both photon energies. The associated DOS can be visualized by examining the momentum-integrated EDCs in Figure 4(c) and Figure 4(d), which exhibit several distinct peaks (black arrows) that arise from flat bands. Figure 4(e) and Figure 4(f) show the momentum-integrated EDCs as a function of photon energy along $\overline{\Gamma}$--$\overline{\text{K}}$--$\overline{\text{M}}$--$\overline{\text{K}}$--$\overline{\Gamma}$ and $\overline{\text{K}}$--$\overline{\Gamma}$--$\overline{\text{K}}$; see Supplementary Note 4 and Supplementary Figures (5-6) for additional data. The peaks related to the four flat bands in these integrated spectra also occur at the four aforementioned energies, as expected (black arrows). A closer examination of the spectra reveals a slight dispersion in the flat bands around $\sim$ - 0.3 eV and $\sim$ - 0.5 eV, which could be attributed to hopping between the nearest and/or next-nearest-neighbor atoms within the same or adjacent kagome planes \cite{YCr6Ge6}.

\indent It is interesting to further consider band dispersions along $\overline{\text{K}}$--$\overline{\Gamma}$--$\overline{\text{K}}$. Results at 50 eV and 60 eV photon energies presented in Figure 5(a,b) show the presence of a linear Dirac-like state formed by the innermost band (dashed blue lines culminating in the blue arrow), which looks like a hole pocket that merges near the E$_{\text{F}}$. For additional related ARPES results, see Supplementary Note 4 and Supplementary Figure 6. The linear Dirac-like band is better resolved at 60 eV photon energy and is identified by symbol $\zeta$ for clarity. The asymmetry seen in the spectra around $\overline{\text{K}}$ is likely due to well-known photoemission cross-section effects, in ARPES measurements. To gain further insight into the nature of the innermost linear Dirac-like band (blue arrow in Figure 5(a.b)), we consider momentum distribution curves obtained by integrating the spectra over a narrow 10 meV binding energy range below the Fermi level. The results obtained from spectra taken at various photon energies are presented in Supplementary Note (4-5) and Supplementary Figure (6-7). The innermost band at the $\overline{\text{K}}$ is seen to show dispersion, indicating its bulk origin, see Supplementary Note 5 and Supplementary Figure 7 for details. The bulk bands plotted in Figure 5(c) also show a linear Dirac-like band around the $\overline{\text{K}}$ point (blue arrow) in reasonable accord with our experimental results. The band characters in Figure 5(c) indicate that the linear Dirac-like state is dominated by the Ti \textit{$d_{x^{2}-y^{2}}$} orbitals.

\indent Interestingly, there are many band inversions in the bulk band structure of Figure 5(c). To unravel the topology of YbTi$_3$Bi$_4$, we have calculated the Z$_2$ invariant by following the evolution of the Wannier charge center and also via the parity based methods, see Supplementary Figure 8. In this way, we found the strong Z$_2$ invariant to vanish ($\nu_0=0$) and the weak invariants to be: $\nu_1=0,\nu_2=1,\nu_3=1$. Figure 5(d) presents a DFT-based projection in which both the bulk and surface bands are included. These results are in line with much of the discussion above of bulk and surface states and their signatures in the experimental ARPES spectra, and exhibit complex hybridization of the bulk and surface states at the $\overline{\text{K}}$ point, see Supplementary Note 6 and Supplementary Figure 9 for further details.\\ 

\indent Defining features of the valence electronic structure of YbTi$_3$Bi$_4$ include: (i) a linear Dirac-like state near E$_{\text{F}}$ at the $\overline{\text{K}}$ point; (ii) based on DFT calculations, a saddle point vHS at $\overline{\text{M}}$ lying about 150 meV above E$_{\text{F}}$, along with other two vHSs about 250 meV below the E$_{\text{F}}$; and (iii) four flat bands spanning across a large region of the BZ, with two of these originating from Yb 4\textit{f} states and the other two from the Ti 3\textit{d} kagome states. These features should be contrasted with the low-energy electronic spectrum of vanadium-based kagome materials such as \textit{A}V$_3$Sb$_5$ in which the physics responsible for producing exotic phenomena, such as superconductivity, is dominated by the interplay of the saddle-point vHSs and a topological surface state, both located at $\overline{\text{M}}$ in close proximity to the E$_{\text{F}}$ \cite{Kun2023AV3Sb5, Takemi2022KV3Sb5, Soohyun2021RbV3Sb5, Ming}. Absence of vHSs near E$_{\text{F}}$ could be the reason for lack of superconductivity and charge ordering in YbTi$_3$Bi$_4$. Flat bands have been predicted theoretically in \textit{A}V$_3$Sb$_5$: one flat band lies about 1 eV above the E$_{\text{F}}$ (inaccessible to ARPES), while the other flat band around 1.2 eV below E$_{\text{F}}$ is not well resolved in ARPES experiments \cite{Ortiz2, Ming}. In contrast, we have clearly shown the presence of several flat bands in YbTi$_3$Bi$_4$. It should be possible to tune the chemical potential in YbTi$_3$Bi$_4$ so that it lies near the vHSs and the flat bands. Magnetism in YbTi$_3$Bi$_4$ can also be tuned by replacing Yb with other rare-earth elements \cite{BrendenLn134}. Our work thus sets the stage for further exploration of YbTi$_3$Bi$_4$, including studies involving its exfoliation down to few monolayers.\\ 

\maketitle  
\begin{center}\textbf{IV. CONCLUSIONS}
\end{center}
\indent In summary, we have synthesized a nonmagnetic Ti-based kagome material (YbTi$_3$Bi$_4$) and analyzed its electronic structure in-depth using a combination of magneto-transport, ARPES, and parallel first-principles calculations. The ARPES measured FS and band dispersions are in good accord with our theoretical predictions. We provide compelling evidence for the presence of multiple flat bands and vHSs in YbTi$_3$Bi$_4$. The flat bands originate from both the Ti-based kagome lattice and the Yb$^{2+}$4\textit{f} orbitals. Our work highlights the importance of the Ti-based $Ln$Ti$_3$Bi$_4$ kagome family as a potential platform for exploring exotic phases.\\

\indent
\textit{Note added.} While the manuscript was under review, several related studies have been published \cite{Chen, Guo134, Chen134, BrendenLn134}.\\ 



\vspace{2ex}

\noindent \textbf{Acknowledgments}\\
\noindent M.N. acknowledges support from the Air Force Office of Scientific Research MURI (Grant No. FA9550-20-1-0322) and the National Science Foundation (NSF) CAREER award DMR-1847962. Work performed by B.R.O. is sponsored by the Laboratory Directed Research and Development Program of Oak Ridge National Laboratory, managed by UT-Battelle, LLC, for the US Department of Energy. DGM acknowledges the support from AFOSR MURI (Novel Light-Matter Interactions in Topologically Non-Trivial Weyl Semimetal Structures and Systems), grant FA9550-20-1-0322. The work at Northeastern University was supported by the Air Force Office of Scientific Research under Award No. FA9550-20-1-0322, and it benefited from the computational resources of Northeastern University’s Advanced Scientific Computation Center (ASCC), the Discovery Cluster, and the Quantum Materials and Sensing Institute (QMSI). This research used resources of the Advanced Light Source, a U.S. Department of Energy Office of Science User Facility, under Contract No. DE-AC02-05CH11231. We thank Sung-Kwan Mo and Jonathan Denlinger for beamline assistance at the Advanced Light Source (ALS), Lawrence Berkeley National Laboratory. We thank Makoto Hashimoto and Donghui Lu for the beamline assistance at SSRL end station 5-2. The use of Stanford Synchrotron Radiation Lightsource (SSRL) in SLAC National Accelerator Laboratory is supported by the U.S. Department of Energy, Office of Science, Office of Basic Energy Sciences under Contract No. DE-AC02-76SF00515.\\



\clearpage
\widetext
\begin{center}
\textbf{\large Supplemental Materials for \\~\\Diverse electronic landscape of the kagome metal YbTi$_3$Bi$_4$}
\end{center}
\setcounter{equation}{0}
\setcounter{figure}{0}
\setcounter{table}{0}
\setcounter{page}{1}
\makeatletter
\renewcommand{\theequation}{S\arabic{equation}}
\renewcommand{\thefigure}{S\arabic{figure}}
\renewcommand{\bibnumfmt}[1]{[#1]}
\renewcommand{\citenumfont}[1]{#1}
\renewcommand{\figurename}{{Supplementary Fig.}}
\renewcommand{\thefigure}{{{\arabic{figure}}}}
\renewcommand{\tablename}{Supplementary Table}
\renewcommand{\thetable}{\arabic{table}}
\def\bibsection{\refname}
\renewcommand{\refname}{\noindent\textbf{Supplementary References}}

\noindent\textbf{Supplementary Note 1. DFT based electronic bands along various high symmetry lines.}\\
Supplementary Figure 1 presents contributions to the valence bands from various Ti \textit{d} orbitals (Ti \textit{$d_{x^{2}-y^{2}}$}, Ti \textit{$d_{z^{2}}$}, Ti \textit{d$_{xy}$}, Ti \textit{d$_{xz}$}, Ti \textit{d$_{yz}$}). These  plots indicate that the flat bands originate from the Ti kagome plane with major contribution from Ti \textit{$d_{x^{2}-y^{2}}$} and Ti \textit{d$_{xy}$} orbitals.\\

\noindent\textbf{Supplementary Note 2. Fermi surface of YbTi$_3$Bi$_4$.}\\
Supplementary Figure 2 presents the Fermi surface (FS) map obtained via ARPES, and the corresponding calculated bulk FS. ARPES measurements use a photon energy of 100 eV. The FS shows a complex metallic band structure. Multiple Brillouin zones are shown. Theoretically predicted FS is seen to be in reasonable accord with the experimental results.\\ 

\begin{figure} 
	\includegraphics[width=15cm]{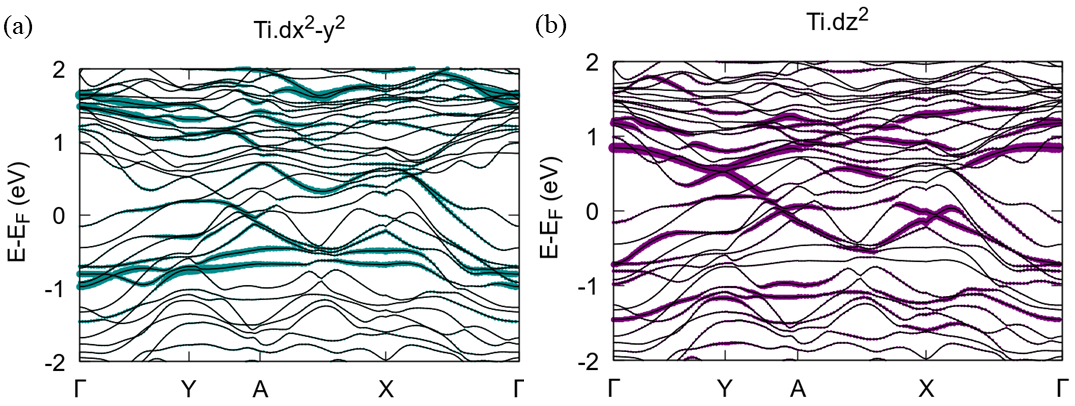} 
\includegraphics[width=15cm]{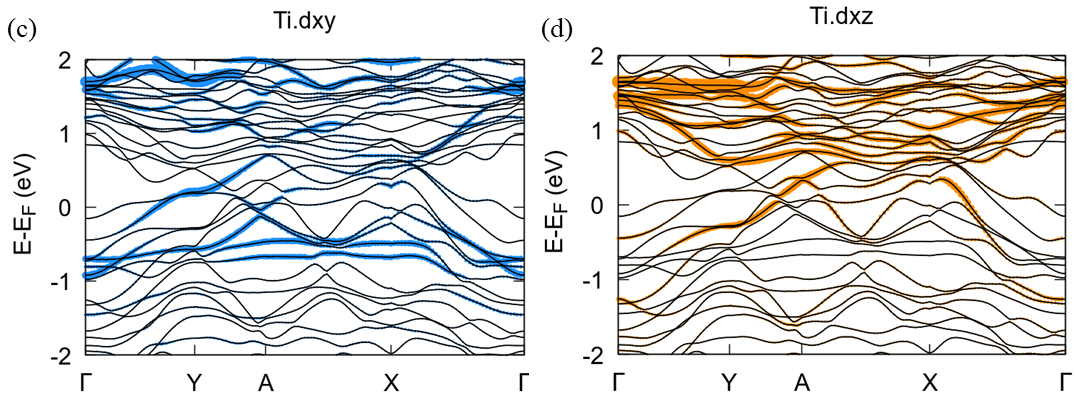} 
\includegraphics[width=7.5cm]{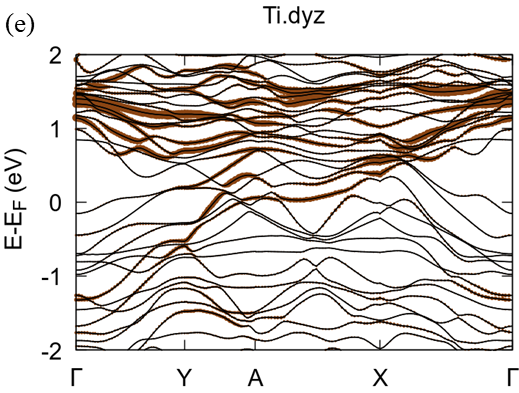}
       \vspace{-1ex}
	\caption{Orbital contributions from various Ti orbitals. Band dispersions along high symmetry lines showing the contributions from various Ti \textit{d} orbitals. Weighted color plots provide relative contributions of various orbitals to the energy bands.} 
\label{fig1}
\end{figure}

\begin{figure} 
	\includegraphics[width=10cm]{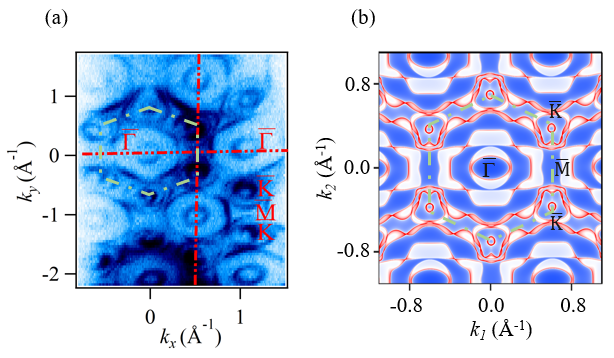}
    	\caption{Fermi surface of YbTi$_3$Bi$_4$. (a) Fermi surface measured using ARPES with a photon energy of 100 eV. (b) Projected bulk Fermi surface obtained using DFT.}
\label{fig2}
\end{figure}

\noindent\textbf{Supplementary Note 3. Observation of two bulk bands at the $\overline{\Gamma}$ point.}\\
Supplementary Figure 3 presents the ARPES-measured band dispersion along the $\overline{\Gamma}$–$\overline{\text{M}}$–$\overline{\Gamma}$–$\overline{\text{M}}$–$\overline{\Gamma}$ high-symmetry line using a photon energy of 75 eV. We clearly observe the presence of two bands around the $\overline{\Gamma}$ point, as seen in Supplementary Figure 3(a). This is better seen in the zoomed-in ARPES spectra shown in Supplementary Figure 3(b). The MDCs, obtained by integrating within 10 meV below the Fermi level, are presented in Supplementary Figure 3(c), along with the fits using two Gaussian peaks  ($\alpha$ and $\beta$) which further confirm the presence of two bulk bands. The bands around the $\overline{\Gamma}$ point are seen to be more intense in the second BZ compared to the first BZ.\\ 

\begin{figure} 
	\includegraphics[width=11cm]{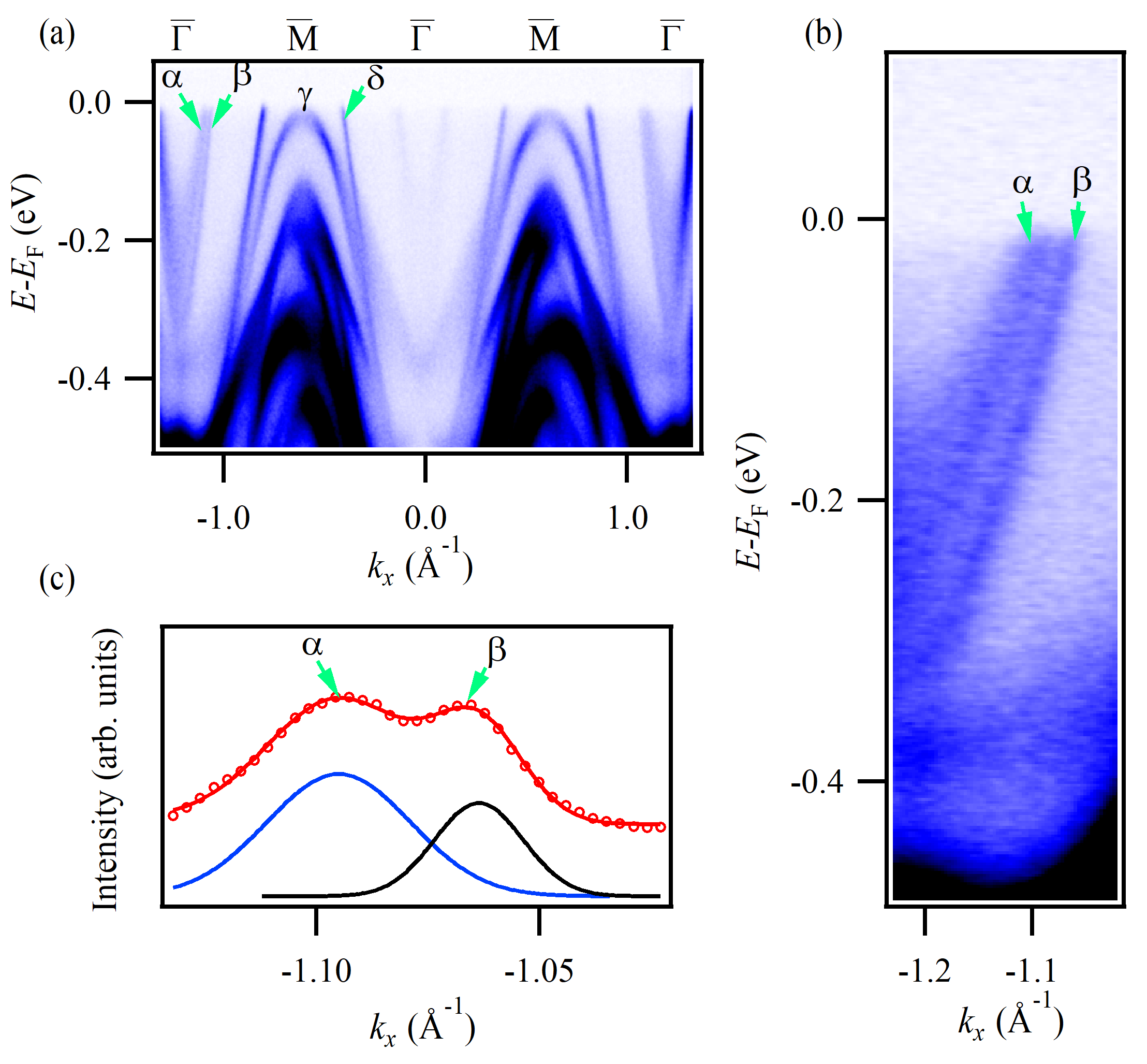} 
	\caption{Observation of two bands around the $\overline{\Gamma}$ point. (a) Experimental band dispersions along the $\overline{\Gamma}$--$\overline{\text{M}}$--$\overline{\Gamma}$--$\overline{\text{M}}$--$\overline{\Gamma}$ high-symmetry line with a photon energy of 75 eV. (b) Zoomed-in view of the band dispersions displaying the two bands ($\alpha$ and $\beta$) around $\overline{\Gamma}$. (c) Two Gaussian peak fits for the momentum distribution curves (MDCs) obtained by integrating the spectrum over an energy window of 10 meV below the Fermi level. Experimental data are represented by red open pentagons, while the fit is given by the red solid-line. The two Gaussians underlying the fit are shown by the blue and black solid lines.}
\label{fig3}
\end{figure}

\begin{figure} 
	\includegraphics[width=14cm]{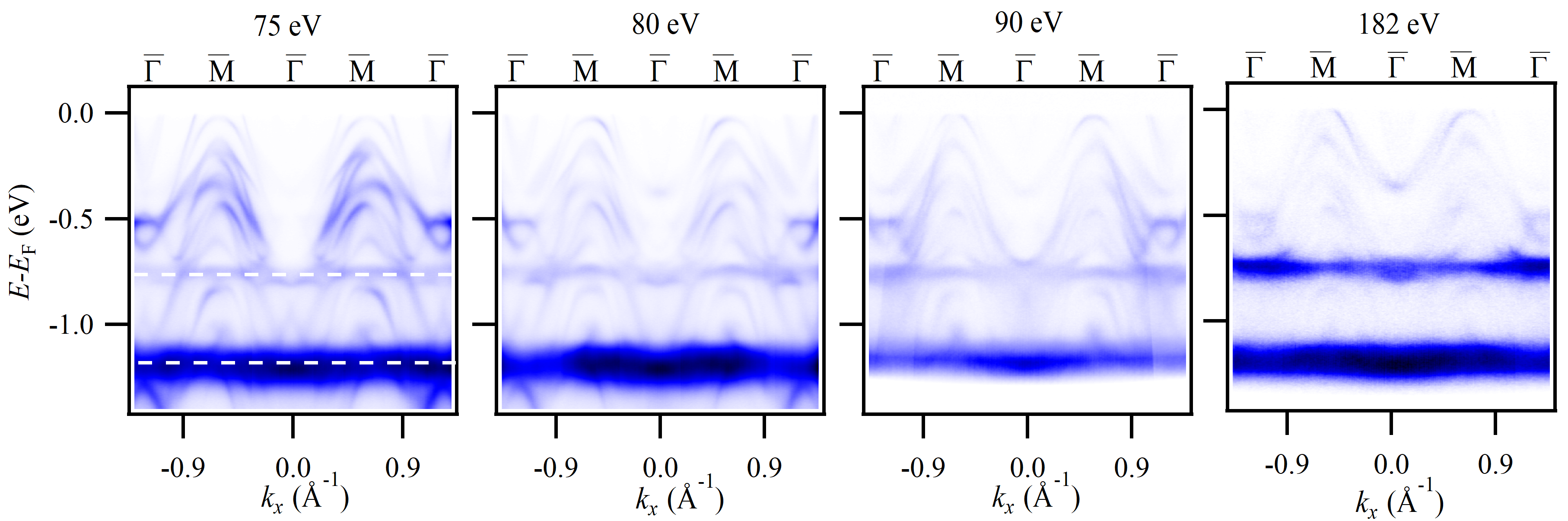} 
       \vspace{-1ex}
	\caption{Observation of Yb 4\textit{f} derived flat bands. ARPES measured band dispersions along the $\overline{\Gamma}$--$\overline{\text{M}}$--$\overline{\Gamma}$--$\overline{\text{M}}$--$\overline{\Gamma}$ line over a wide energy range using various photon energies (noted on the panels).}
\label{fig3}
\end{figure}

\noindent\textbf{Supplementary Note 4. Observation of flat bands along various high-symmetry lines}\\
Supplementary Figures. 4-6 presents the band-dispersions obtained via ARPES along various high-symmetry lines. In Supplementary Figure 4, the band dispersions along the $\overline{\Gamma}$--$\overline{\text{M}}$--$\overline{\Gamma}$--$\overline{\text{M}}$--$\overline{\Gamma}$ line over a wide energy range is presented. Measurements involved several different photon energies. These plots allow us to assign the flat bands at -0.7 eV and -1.1 eV to Yb 4\textit{f} states. A suppression of photoemission intensity from the bands which do not possess Yb 4\textit{f} states and an enhancement of photoemission intensity from Yb 4\textit{f} states at -0.7 eV and -1.1 eV is also visible at 182 eV photon energy (on resonance) compared to all other photon energies. Band dispersions along $\overline{\Gamma}$--$\overline{\text{K}}$--$\overline{\text{M}}$--$\overline{\text{K}}$--$\overline{\Gamma}$ are also presented in Supplementary Figure 5 using various photon energies. The measured band dispersions as well as the integrated energy distribution curves (EDCs) show the presence of four bands (highlighted by black arrows), which lie at -0.3 eV, -0.5 eV, -0.7 eV, and -1.1 eV.

Supplementary Figure 6 illustrates the band dispersions along the $\overline{\text{K}}$–$\overline{\Gamma}$–$\overline{\text{K}}$ line obtained by using various photon energies ranging from 40-60 eV. The black arrow and dotted line in Supplementary Figure 6(a) highlight the linear Dirac-like band at the $\overline{\text{K}}$ point, while the solid black lines mark the peaks identified from the MDCs at a binding energy of 10 meV below the Fermi level. As seen in Supplementary Figure 6, MDC cuts taken around 10 meV below the Fermi level are expected to reveal two peaks each from the linear Dirac-like state at the $\overline{\text{K}}$ point. For clarity, integrated MDCs and peak positions for the $\zeta$ band are marked by black lines in the figure. At photon energies of 40 eV and 45 eV, a single peak is observed for each $\overline{\text{K}}$ point, whereas higher photon energies reveal two distinct peaks, indicated by the two black solid lines. The absence of two peaks at certain photon energies may be due to photoemission cross-section effects, which can suppress spectral intensity in ARPES measurements.\\

\noindent\textbf{Supplementary Note 5. Dispersion of the inner-most band at the $\overline{\text{K}}$ point.}\\
To track the dispersion of the innermost band (labeled $\zeta$, we fit the MDC at the 10 meV binding energy along the $\overline{\text{K}}$–$\overline{\Gamma}$–$\overline{\text{K}}$ line. Our focus is on the innermost peak ($\zeta$), highlighted by the black arrow in Supplementary Figure 6(a), with the dashed black line added for better visualization. Supplementary Figure 7(a) presents ARPES cuts and MDC fits for a photon energy of 40 eV along the $\overline{\text{K}}$–$\overline{\Gamma}$–$\overline{\text{K}}$ line, and it reveals a single peak along the $\pm$k$_y$ direction for photon energies of 40 eV and 45 eV (blue curves). The peaks are located at $\pm$0.57 \AA$^{-1}$ and $\pm$0.6 \AA$^{-1}$ for photon energies of 40 eV and 45 eV, respectively. The observed asymmetry in the spectra around the $\pm$k$_y$ directions near $\overline{\text{K}}$ is likely due to photoemission cross-section effects. When the photon energy is increased to 50 eV, two peaks become apparent at $\pm$0.77 \AA$^{-1}$ and $\pm$0.58 \AA$^{-1}$. At 55 eV, the spectra resemble those at 50 eV, with the innermost peaks observed at $\pm$0.77 \AA$^{-1}$ and $\pm$0.55 \AA$^{-1}$. For a photon energy of 58 eV, the innermost peaks shift to $\pm$0.87 \AA$^{-1}$ and $\pm$0.6 \AA$^{-1}$. Finally, at 60 eV, the peaks are observed at $\pm$0.84 \AA$^{-1}$ and $\pm$0.62 \AA$^{-1}$. This detailed MDC fitting around the $\overline{\text{K}}$ points confirms the dispersion of the innermost peak, $\zeta$, consistent with the analysis in the main manuscript.\\

\begin{figure} 
	\includegraphics[width=8cm]{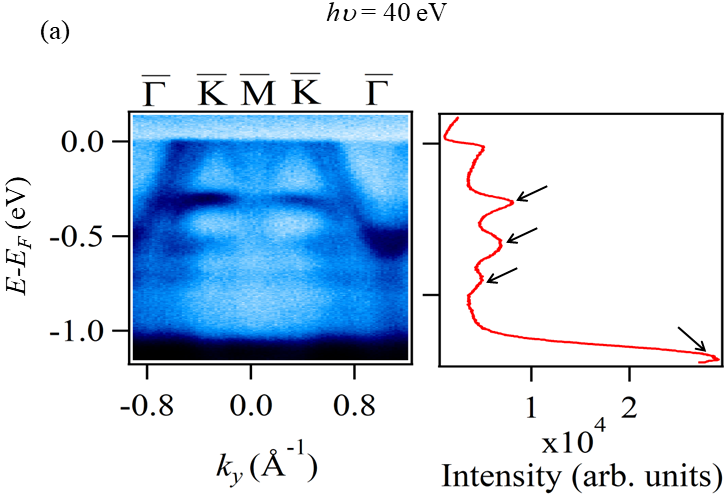} 
     \includegraphics[width=8cm]{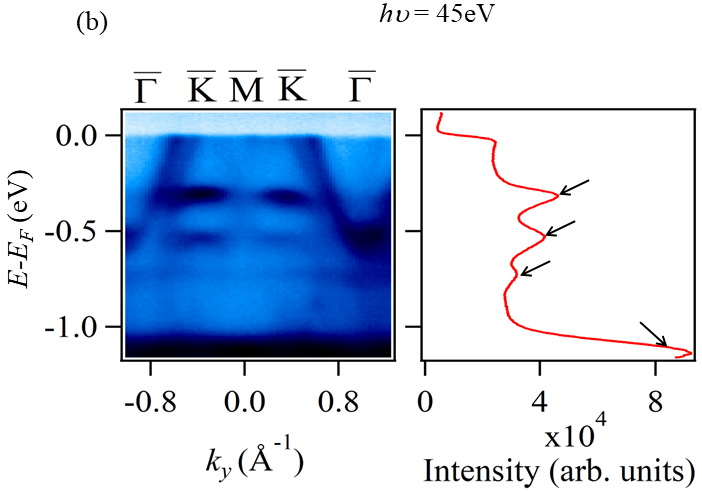} 
\includegraphics[width=8cm]{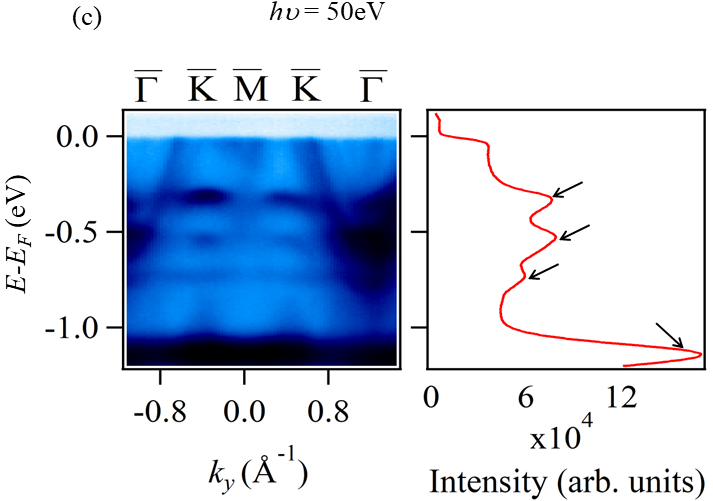}
 	\caption{Flat bands along the $\overline{\Gamma}$--$\overline{\text{K}}$--$\overline{\text{M}}$--$\overline{\text{K}}$--$\overline{\Gamma}$ high-symmetry line. Experimental band dispersions and integrated energy distribution curves (EDCs) over the entire momentum range along the $\overline{\Gamma}$--$\overline{\text{K}}$--$\overline{\text{M}}$--$\overline{\text{K}}$--$\overline{\Gamma}$ line obtained by using a photon energy of (a) 40 eV, (b) 45 eV, (c) 50 eV. EDC peaks (black arrows) correspond to flat bands at $\sim$ -0.3 eV, $\sim$ -0.5 eV, $\sim$ -0.7 eV with the most intense peak at around $\sim$ -1.1 eV.}
\label{fig4}
\end{figure}

\noindent\textbf{Supplementary Note 6. Bulk DFT calculations projected along the $\overline{\text{K}}$--$\overline{\Gamma}$--$\overline{\text{K}}$ high-symmetry line.}\\
Supplementary Figure 9 presents the projected bulk band structure along $\overline{\text{K}}$--$\overline{\Gamma}$--$\overline{\text{K}}$ line. Presence of flat bands and a linear Dirac-like state at the $\overline{\text{K}}$ point can be seen (red arrow).

\begin{figure} 
	\includegraphics[width=12cm]{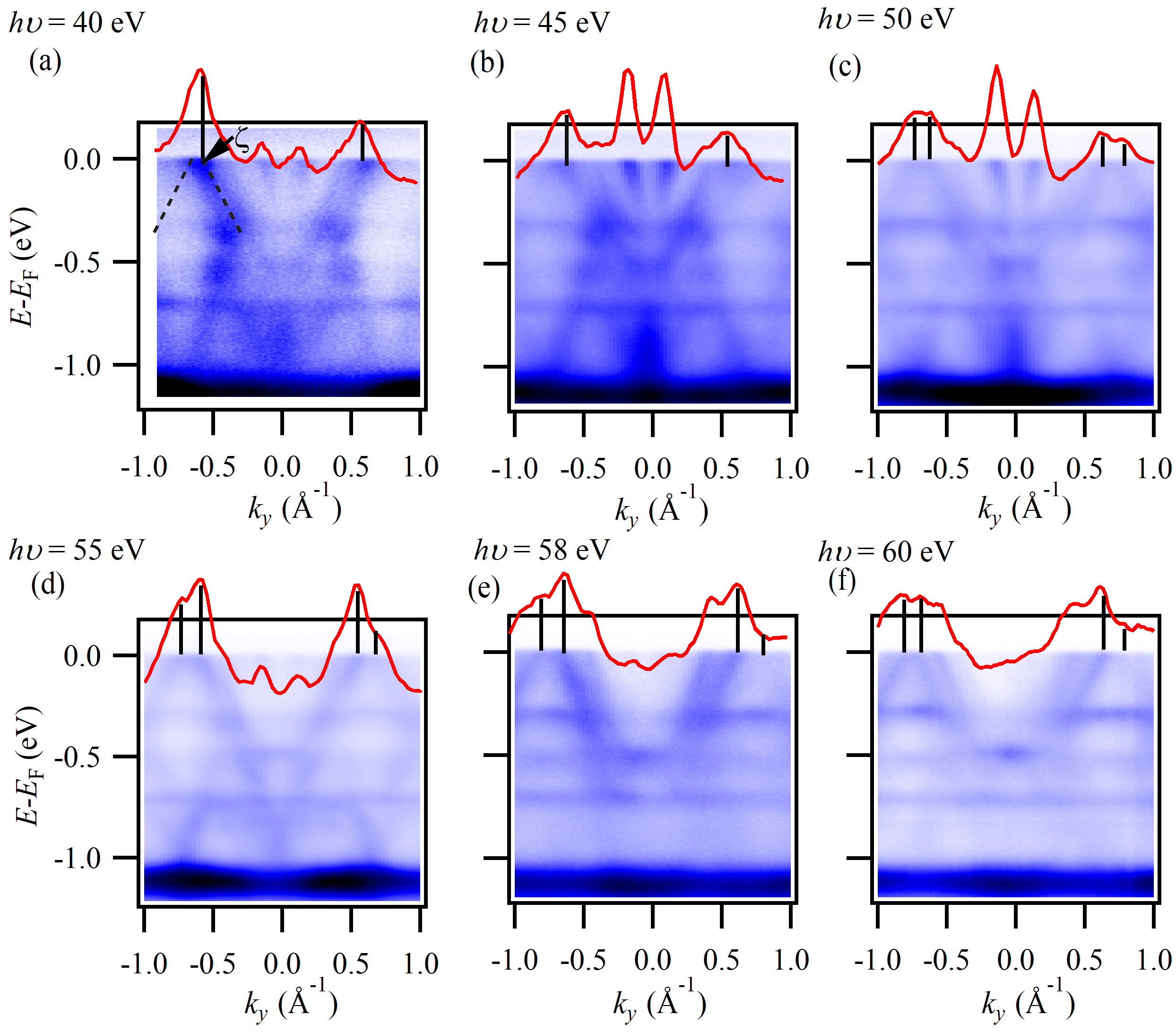}
	\caption{Dirac-like state at the $\overline{\text{K}}$ point. Experimental band dispersions along $\overline{\text{K}}$--$\overline{\Gamma}$--$\overline{\text{K}}$ obtained by using photon energies of (a) 40 eV, (b) 45 eV, (c) 50 eV, (d) 55 eV, (e) 58 eV, and (f) 60 eV. The integrated MDCs within 10 meV below the Fermi level for each spectrum are overlaid to enhance the visualization of the peaks formed by the innermost band ($\zeta$).}
\label{fig2}
\end{figure}

\begin{figure} 
	\includegraphics[width=12cm]{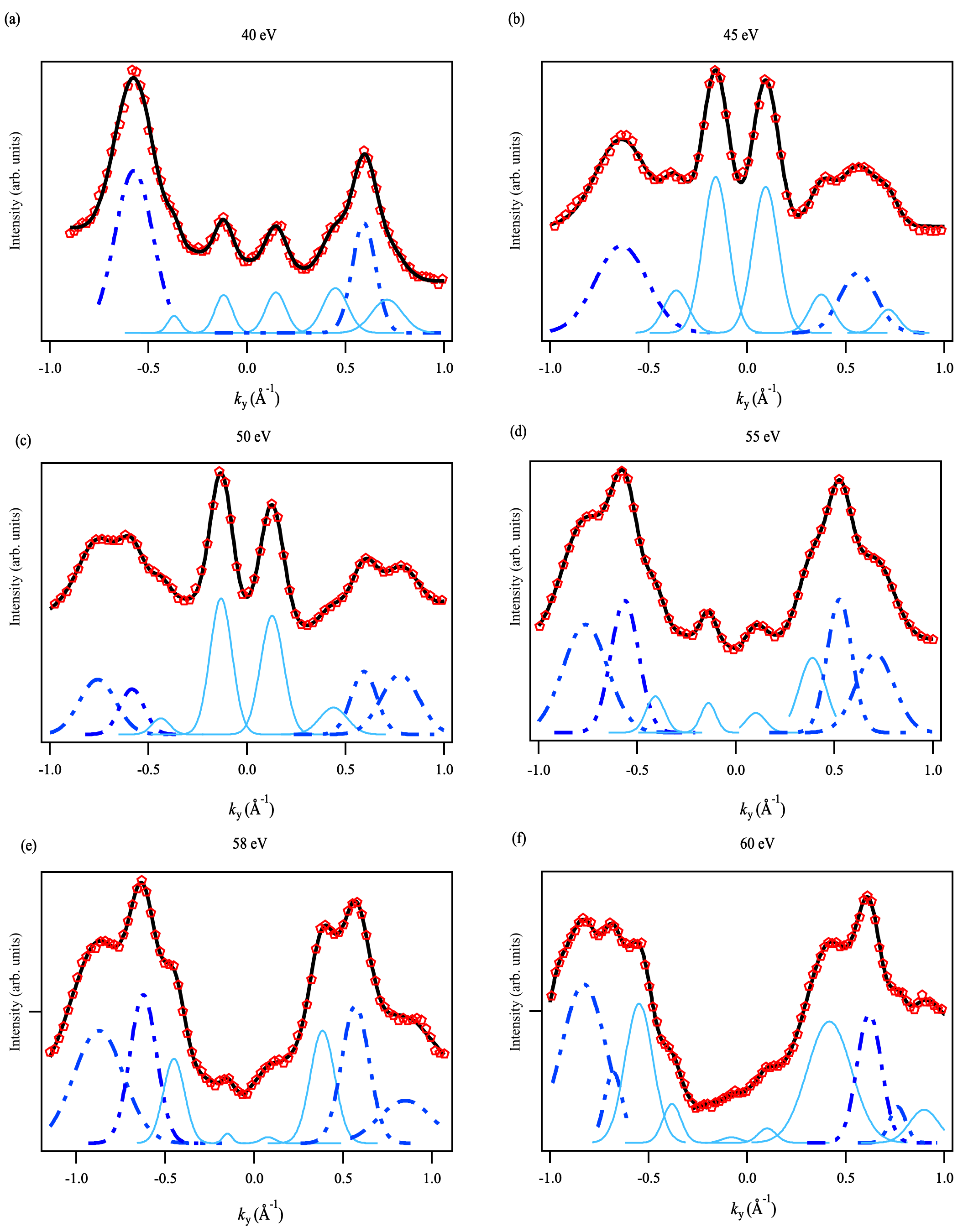}
  	\caption{Tracking the dispersion of the innermost band $\zeta$ at the $\overline{\text{K}}$ point. The figure presents multiple-Gaussian-peak fit of the MDCs along the $\overline{\text{K}}$--$\overline{\Gamma}$--$\overline{\text{K}}$ line, along with the integrated MDCs within 10 meV below the Fermi level. Experimental data are represented by red open pentagons, while the fitted spectrum is given by the black line. The two underlying Gaussians used to fit the data are shown by the cyan-solid and blue-dotted lines. Peaks associated with the innermost band $\zeta$ are highlighted by the blue dot-dashed lines. Data were collected at different photon energies as indicated in the plots.}
\label{fig2}
\end{figure}


\begin{figure} 
	\includegraphics[width=6cm]{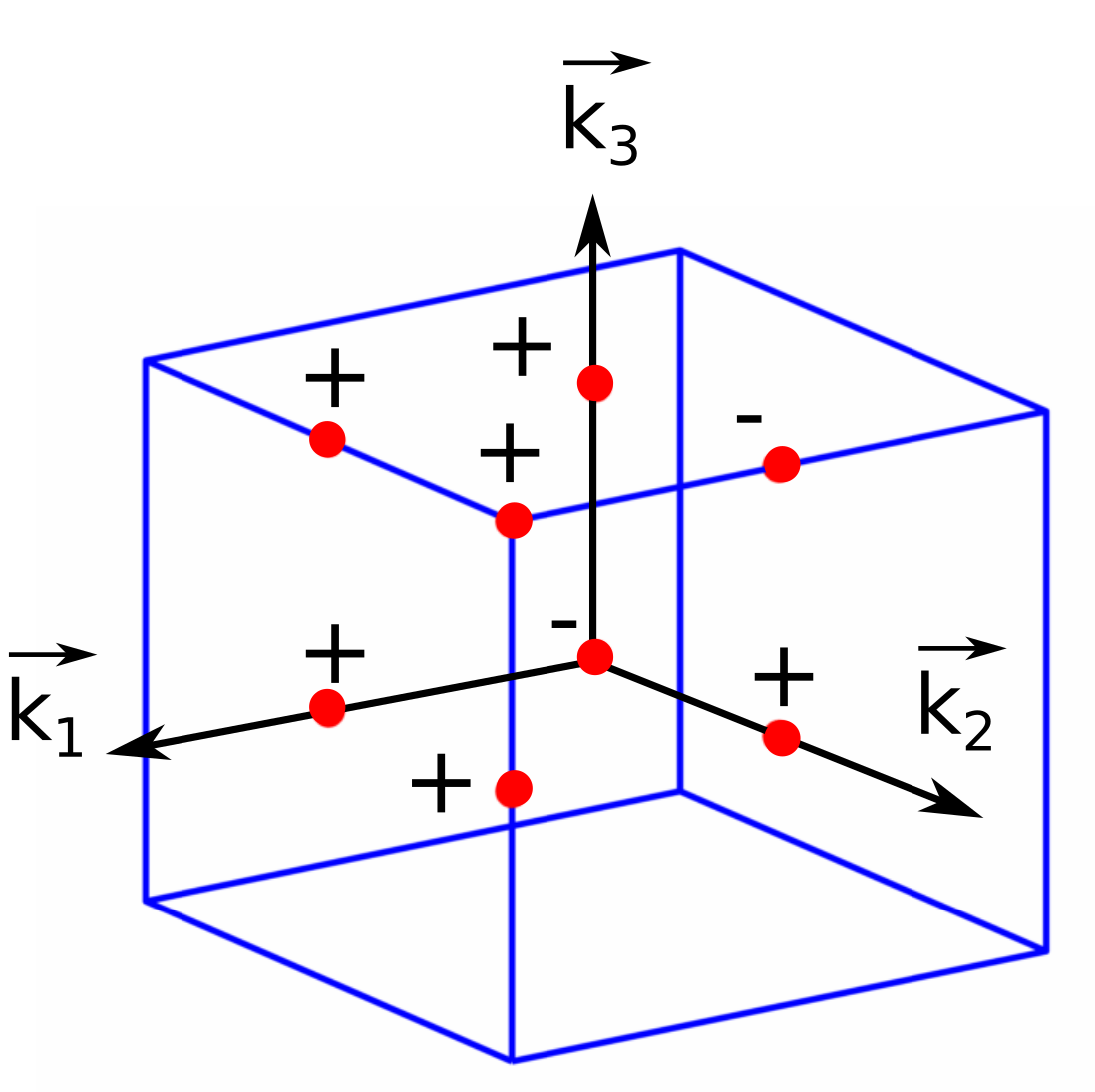} 
      \caption{Product of the parities of the occupied bands at various TRIM points. The calculated Z$_2$ invariant is ($\nu_0=0,\nu_1=0,\nu_2=1,\nu_3=1)$.}
\label{fig3}
\end{figure}

\begin{figure}
\centering 
	\includegraphics[width=6cm]{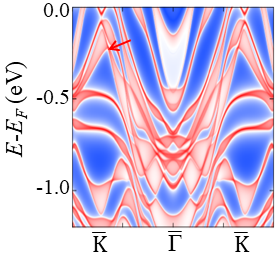} 
	\caption{DFT based projected bulk band dispersions. Band dispersions along the $\overline{\text{K}}$--$\overline{\Gamma}$--$\overline{\text{K}}$ high-symmetry line.}
\label{fig6}
\end{figure}

\begin{thebibliography}{60}
\bibitem{Syozi}I. Syôzi, \href{https://doi.org/10.1143/ptp/6.3.306} {Prog. Theor. Phys. \textbf{6}, 306 (1951)}.
\bibitem{Zhou}Y. Zhou, K. Kanoda, and T.-K. Ng, \href{https://doi.org/10.1103/RevModPhys.89.025003} {Rev. Mod. Phys. \textbf{89}, 025003 (2017)}.
\bibitem{Neupert}T. Neupert, M. M. Denner, J.-X. Yin, R. Thomale, and M. Z. Hasan, \href{https://doi.org/10.1038/s41567-021-01404-y} {Nat. Phys. \textbf{18}, 137 (2022)}.
\bibitem{Guo}H.-M. Guo and M. Franz, \href{https://doi.org/10.1103/PhysRevB.80.113102} {Phys. Rev. B \textbf{80}, 113102 (2009)}.
\bibitem{Tang}E. Tang, J.-W. Mei, and X.-G. Wen, \href{https://doi.org/10.1103/PhysRevLett.106.236802} {Phys. Rev. Lett. \textbf{106}, 236802 (2011)}.
\bibitem{Han}T.-H. Han, J. S. Helton, S. Chu, D. G. Nocera, J. A. Rodriguez-Rivera, C. Broholm, and Y. S. Lee, \href{https://doi.org/10.1038/nature11659} {Nature \textbf{492}, 406 (2012)}.
\bibitem{Ye}L. Ye, M. Kang, J. Liu, F. von Cube, C. R. Wicker, T. Suzuki, C. Jozwiak, A. Bostwick, E. Rotenberg, D. C. Bell, L. Fu, R. Comin, and J. G. Checkelsky, \href{https://doi.org/10.1038/nature25987} {Nature \textbf{555}, 638 (2018)}.
\bibitem{Lin}Z. Lin, J.-H. Choi, Q. Zhang, W. Qin, S. Yi, P. Wang, L. Li, Y. Wang, H. Zhang, and Z. Sun, \href{https://doi.org/10.1103/PhysRevLett.121.096401} {Phys. Rev. Lett. \textbf{121}, 096401 (2018)}.
\bibitem{Yin1}J.-X. Yin, S. S. Zhang, G. Chang, Q. Wang, S. S. Tsirkin, Z. Guguchia, B. Lian, H. Zhou, K. Jiang, I. Belopolski, N. Shumiya, D. Multer, M. Litskevich, T. A. Cochran, H. Lin, Z. Wang, T. Neupert, S. Jia, H. Lei, and M. Z. Hasan, \href{https://doi.org/10.1038/s41567-019-0426-7} {Nat. Phys. \textbf{15}, 443 (2019)}.
\bibitem{Yin2}J.-X. Yin, W. Ma, T. A. Cochran, X. Xu, S. S. Zhang, H.-J. Tien, N. Shumiya, G. Cheng, K. Jiang, B. Lian, Z. Song, G. Chang, I. Belopolski, D. Multer, M. Litskevich, Z.-J. Cheng, X. P. Yang, B. Swidler, H. Zhou, H. Lin, T. Neupert, Z. Wang, N. Yao, T.-R. Chang, S. Jia, and M. Z. Hasan, \href{https://doi.org/10.1038/s41586-020-2482-7} {Nature \textbf{583}, 533 (2020)}.
\bibitem{Kang1}M. Kang, L. Ye, S. Fang, J.-S. You, A. Levitan, M. Han, J. I. Facio, C. Jozwiak, A. Bostwick, and E. Rotenberg, \href{https://doi.org/10.1038/s41563-019-0531-0} {Nat. Mater. \textbf{19}, 163 (2020)}.
\bibitem{Kang2}M. Kang, S. Fang, L. Ye, H. C. Po, J. Denlinger, C. Jozwiak, A. Bostwick, E. Rotenberg, E. Kaxiras, and J. G. Checkelsky, \href{https://doi.org/10.1038/s41467-020-17465-1} {Nat. Commun. \textbf{11}, 4004 (2020)}.
\bibitem{Nirmal}N. J. Ghimire and I. I. Mazin, \href{https://doi.org/10.1038/s41563-019-0589-8} {Nat. Mater. \textbf{19}, 137 (2020)}.
\bibitem{Li}M. Li, Q. Wang, G. Wang, Z. Yuan, W. Song, R. Lou, Z. Liu, Y. Huang, Z. Liu, H. Lei, Z. Yin, and S. Wang, \href{https://doi.org/10.1038/s41467-021-23536-8} {Nat. Commun. \textbf{12}, 3129 (2021)}.
\bibitem{Regmi}S. Regmi, T. Fernando, Y. Zhao, A. P. Sakhya, G. Dhakal, I. Bin Elius, H. Vazquez, J. D. Denlinger, J. Yang, and J.-H. Chu, \href{https://doi.org/10.1038/s43246-022-00318-3} {Commun. Mater. \textbf{3}, 100 (2022)}.
\bibitem{Yinkagome}J.-X. Yin, B. Lian, and M. Z. Hasan, \href{https://doi.org/10.1038/s41586-022-05516-0} {Nature \textbf{612}, 647 (2022)}.
\bibitem{Balents}L. Balents, \href{https://doi.org/10.1038/nature08917} {Nature \textbf{464}, 199 (2010)}.
\bibitem{Yang}J. Yang, X. Yi, Z. Zhao, Y. Xie, T. Miao, H. Luo, H. Chen, B. Liang, W. Zhu, Y. Ye, J.-Y. You, B. Gu, S. Zhang, F. Zhang, F. Yang, Z. Wang, Q. Peng, H. Mao, G. Liu, Z. Xu, H. Chen, H. Yang, G. Su, H. Gao, L. Zhao, and X. J. Zhou, \href{https://doi.org:10.1038/s41467-023-39620-0} {Nat. Commun. \textbf{14}, 4089 (2023)}. 
\bibitem{Ortiz1}B. R. Ortiz, L. C. Gomes, J. R. Morey, M. Winiarski, M. Bordelon, J. S. Mangum, I. W. Oswald, J. A. Rodriguez-Rivera, J. R. Neilson, and S. D. Wilson, \href{https://doi.org/10.1103/PhysRevMaterials.3.094407} {Phys. Rev. Mater. \textbf{3}, 094407 (2019)}.
\bibitem{Ortiz2}B. R. Ortiz, S. M. Teicher, Y. Hu, J. L. Zuo, P. M. Sarte, E. C. Schueller, A. M. Abeykoon, M. J. Krogstad, S. Rosenkranz, and R. Osborn, \href{https://doi.org/10.1103/PhysRevLett.125.247002} {Phys. Rev. Lett. \textbf{125}, 247002 (2020)}.
\bibitem{Ortiz3}B. R. Ortiz, P. M. Sarte, E. M. Kenney, M. J. Graf, S. M. Teicher, R. Seshadri, and S. D. Wilson, \href{https://doi.org/10.1103/PhysRevMaterials.5.034801} {Phys. Rev. Mater. \textbf{5}, 034801 (2021)}.
\bibitem{Ortiz4}B. R. Ortiz, S. M. L. Teicher, L. Kautzsch, P. M. Sarte, N. Ratcliff, J. Harter, J. P. C. Ruff, R. Seshadri, and S. D. Wilson, \href{https://doi.org/10.1103/PhysRevX.11.041030} {Phys. Rev. X. \textbf{11}, 041030 (2021)}.
\bibitem{Zhao}H. Zhao, H. Li, B. R. Ortiz, S. M. L. Teicher, T. Park, M. Ye, Z. Wang, L. Balents, S. D. Wilson, and I. Zeljkovic, \href{https://doi.org/10.1038/s41586-021-03946-w} {Nature \textbf{599}, 216 (2021)}.
\bibitem{Hu}Y. Hu, X. Wu, B. R. Ortiz, X. Han, N. C. Plumb, S. D. Wilson, A. P. Schnyder, and M. Shi, \href{https://doi.org/10.1103/PhysRevB.106.L241106} {Phys. Rev. B \textbf{106}, L241106 (2022)}.
\bibitem{Jiang}Y.-X. Jiang, J.-X. Yin, M. M. Denner, N. Shumiya, B. R. Ortiz, G. Xu, Z. Guguchia, J. He, M. S. Hossain, X. Liu, J. Ruff, L. Kautzsch, S. S. Zhang, G. Chang, I. Belopolski, Q. Zhang, T. A. Cochran, D. Multer, M. Litskevich, Z.-J. Cheng, X. P. Yang, Z. Wang, R. Thomale, T. Neupert, S. D. Wilson, and M. Z. Hasan, \href{https://doi.org/10.1038/s41563-021-01034-y} {Nat. Mater. \textbf{20}, 1353 (2021)}.
\bibitem{Hao}H. Zhao, H. Li, B. R. Ortiz, S. M. L. Teicher, T. Park, M. Ye, Z. Wang, L. Balents, S. D. Wilson, and I. Zeljkovic, \href{https://doi.org/10.1103/PhysRevB.106.L081101} {Phys. Rev. B \textbf{106}, L081101 (2022)}.
\bibitem{LeiRb135}Q. Yin, Z. Tu, C. Gong, Y. Fu, S. Yan, and H. Lei, \href{10.1088/0256-307X/38/3/037403} {Chin. Phys. Lett. \textbf{38}, 037403 (2021)}.
\bibitem{LuoK135}H. Luo, Q. Gao, H. Liu, Y. Gu, D. Wu, C. Yi, J. Jia, S. Wu, X. Luo, Y. Xu, L. Zhao, Q. Wang, H. Mao, G. Liu, Z. Zhu, Y. Shi, K. Jiang, J. Hu, Z. Xu, and X. J. Zhou, \href{https://doi.org/10.1038/s41467-021-27946-6} {Nat. Commun. \textbf{13}, 273 (2022)}.
\bibitem{Wilson} S. D. Wilson and B. R. Ortiz, \href{https://doi.org/10.1038/s41578-024-00677-y} {Nat. Rev. Mater. \textbf{9}, 420 (2024)}. 
 Nat. Rev. Mater., 9, 420,  (2024).
\bibitem{Nakatsuji}S. Nakatsuji, N. Kiyohara, and T. Higo, \href{https://doi.org/10.1038/nature15723} {Nature \textbf{527}, 212 (2015)}.
\bibitem{Nayak}A. K. Nayak, J. E. Fischer, Y. Sun, B. Yan, J. Karel, A. C. Komarek, C. Shekhar, N. Kumar, W. Schnelle, and J. Kübler, \href{DOI: 10.1126/sciadv.1501870} {Sci. Adv. \textbf{2}, e1501870 (2016)}.
\bibitem{Liu}E. Liu, Y. Sun, N. Kumar, L. Muechler, A. Sun, L. Jiao, S.-Y. Yang, D. Liu, A. Liang, and Q. Xu, \href{https://doi.org/10.1038/s41567-018-0234-5} {Nat. Phys. \textbf{14}, 1125 (2018)}.
\bibitem{Chenweyl}D. F. Liu, A. J. Liang, E. K. Liu, Q. N. Xu, Y. W. Li, C. Chen, D. Pei, W. J. Shi, S. K. Mo, P. Dudin, T. Kim, C. Cacho, G. Li, Y. Sun, L. X. Yang, Z. K. Liu, S. S. P. Parkin, C. Felser, and Y. L. Chen, \href{https://doi.org:doi:10.1126/science.aav2873} {Science \textbf{365}, 1282 (2019)}.
\bibitem{Ma}W. Ma, X. Xu, J.-X. Yin, H. Yang, H. Zhou, Z.-J. Cheng, Y. Huang, Z. Qu, F. Wang, M. Z. Hasan, and S. Jia, \href{https://doi.org/10.1103/ PhysRevLett.126.246602.} {Phys. Rev. Lett. \textbf{126}, 246602 (2021)}.
\bibitem{Wang}R. S. Li, T. Zhang, W. Ma, S. X. Xu, Q. Wu, L. Yue, S. J. Zhang, Q. M. Liu, Z. X. Wang, T. C. Hu, X. Y. Zhou, D. Wu, T. Dong, S. Jia, H. Weng, and N. L. Wang, \href{https://doi.org/10.1103/PhysRevB.107.045115.} {Phys. Rev. B \textbf{107}, 045115 (2023)}.
\bibitem{Gu} X. Gu, C. Chen, W. S. Wei, L. L. Gao, J. Y. Liu, X. Du, D. Pei, J. S. Zhou, R. Z. Xu, Z. X. Yin, W. X. Zhao, Y. D. Li, C. Jozwiak, A. Bostwick, E. Rotenberg, D. Backes, L. S. I. Veiga, S. Dhesi, T. Hesjedal, G. van der Laan, H. F. Du, W. J. Jiang, Y. P. Qi, G. Li, W. J. Shi, Z. K. Liu, Y. L. Chen, and L. X. Yang, \href{https://doi.org/10.1103/PhysRevB.105.155108} {Phys. Rev. B \textbf{105}, 155108 (2022)}.
\bibitem{Asaba}T. Asaba, S. M. Thomas, M. Curtis, J. D. Thompson, E. D. Bauer, and F. Ronning, \href{https://doi.org/10.1103/PhysRevB.101.174415} {Phys. Rev. B \textbf{101}, 174415 (2020)}.
\bibitem{Zeng}H. Zeng, G. Yu, X. Luo, C. Chen, C. Fang, S. Ma, Z. Mo, J. Shen, M. Yuan, and Z. Zhong, \href{https://doi.org/10.1016/j.jallcom.2021.163356} {J. Alloys Compd.. \textbf{899}, 163356 (2022)}.
\bibitem{Y16}N. J. Ghimire, R. L. Dally, L. Poudel, D. C. Jones, D. Michel, N. T. Magar, M. Bleuel, M. A. McGuire, J. S. Jiang, J. F. Mitchell, J. W. Lynn, and I. I. Mazin, \href{DOI: 10.1126/sciadv.abe2680} {Sci. Adv. \textbf{6}, eabe2680 (2020)}.
\bibitem{Dhakal}G. Dhakal, F. Cheenicode Kabeer, A. K. Pathak, F. Kabir, N. Poudel, R. Filippone, J. Casey, A. P. Sakhya, S. Regmi, C. Sims, K. Dimitri, P. Manfrinetti, K. Gofryk, P. M. Oppeneer, and M. Neupane, \href{https://doi.org:10.1103/PhysRevB.104.L161115} {Phys. Rev. B \textbf{104}, L161115 (2021)}.
\bibitem{Wangymsn6}Q. Wang, K. J. Neubauer, C. Duan, Q. Yin, S. Fujitsu, H. Hosono, F. Ye, R. Zhang, S. Chi, K. Krycka, H. Lei, and P. Dai, \href{https://doi.org:10.1103/PhysRevB.103.014416} {Phys. Rev. B \textbf{103}, 014416 (2021)}.
\bibitem{Kabir}F. Kabir, R. Filippone, G. Dhakal, Y. Lee, N. Poudel, J. Casey, A. P. Sakhya, S. Regmi, R. Smith, P. Manfrinetti, L. Ke, K. Gofryk, M. Neupane, and A. K. Pathak, \href{https://doi.org:10.1103/PhysRevMaterials.6.064404} {Phys. Rev. Mater. \textbf{6}, 064404 (2022)}.
\bibitem{Lv}B. Lv, R. Zhong, X. Luo, S. Ma, C. Chen, S. Wang, Q. Luo, F. Gao, C. Fang, W. Ren, and Z. Zhong, \href{https://doi.org/10.1016/j.jallcom.2023.170356} {J. Alloys Compd.. \textbf{957}, 170356 (2023)}.
\bibitem{Felser}D. Chen, C. Le, C. Fu, H. Lin, W. Schnelle, Y. Sun, and C. Felser, \href{https://doi.org:10.1103/PhysRevB.103.144410} {Phys. Rev. B \textbf{103}, 144410 (2021)}.
\bibitem{Ortiz5}B. R. Ortiz, G. Pokharel, M. Gundayao, H. Li, F. Kaboudvand, L. Kautzsch, S. Sarker, J. P. C. Ruff, T. Hogan, S. J. G. Alvarado, P. M. Sarte, G. Wu, T. Braden, R. Seshadri, E. S. Toberer, I. Zeljkovic, and S. D. Wilson, \href{https://doi.org/10.1103/PhysRevMaterials.7.064201} {Phys. Rev. Mater. \textbf{7}, 064201 (2023)}.
\bibitem{alexander}A. Ovchinnikov and S. Bobev, \href{ https://doi.org/10.1002/ejic.201701426} {Eur. J. Inorg. Chem. \textbf{2018}, 1266 (2018)}.
\bibitem{BrendenLn134}B. R. Ortiz, H. Miao, F. Yang, E. M. Clements, D. S. Parker, J. Yan, A. F. May, and M. A. McGuire, \href{https://doi.org/10.1021/acs.chemmater.3c02289} {Chem. Mater. \textbf{35}, 9756 (2023)}.
\bibitem{Canfield}P. C. Canfield, T. Kong, U. S. Kaluarachchi, and N. H. Jo, \href{https://doi.org:10.1080/14786435.2015.1122248
} {Philos. Mag. \textbf{96}, 84 (2016)}.
\bibitem{Kohn}W. Kohn and L. J. Sham, \href{https://doi.org:10.1103/PhysRev.140.A1133} {Phys. Rev. \textbf{140}, A1133 (1965)}.
\bibitem{kresse1996efficient} G. Kresse and J. Furthm\"{u}ller, \href{https://doi.org/10.1103/PhysRevB.54.11169} {Phys. Rev. B \textbf{54}, 11169 (1996)}.
\bibitem{kresse1999ultrasoft} G. Kresse and D. Joubert, \href{https://doi.org/10.1103/PhysRevB.59.1758} {Phys. Rev. B \textbf{59}, 1758 (1999)}.
\bibitem{SunPRL}J. Sun, A. Ruzsinszky, and J. P. Perdew, \href{https://doi.org:10.1103/PhysRevLett.115.036402} {Phys. Rev. Lett. \textbf{115}, 036402 (2015)}.
\bibitem{SunNature}J. Sun, R. C. Remsing, Y. Zhang, Z. Sun, A. Ruzsinszky, H. Peng, Z. Yang, A. Paul, U. Waghmare, X. Wu, M. L. Klein, and J. P. Perdew, \href{https://doi.org:10.1038/nchem.2535} {Nat. Chem \textbf{8}, 831 (2016)}.

\bibitem{AB1} A. Bansil and M. Lindroos, \href{https://doi.org/10.1103/PhysRevLett.83.5154} {Phys. Rev. Lett., \textbf{83}, 5154, (1999)}.
\bibitem{AB2} S. Sahrakorpi, M. Lindroos, R. S. Markiewicz, and A. Bansil, \href{https://doi.org/10.1103/PhysRevLett.95.157601} {Phys. Rev. Lett., \textbf{95}, 157601, (2005)}.
\bibitem{Sakhyasmbi}A. P. Sakhya, S. Kumar, A. Pramanik, R. P. Pandeya, R. Verma, B. Singh, S. Datta, S. Sasmal, R. Mondal, E. F. Schwier et al., \href{https://doi.org/10.1103/PhysRevB.106.085132} {Phys. Rev. B \textbf{106}, 085132 (2022)}.
\bibitem{anupndsb}A. P. Sakhya, B. Wang, F. Kabir, C.-Y. Huang, M. M. Hosen, B. Singh, S. Regmi, G. Dhakal, K. Dimitri, M. Sprague et al., \href{https://doi.org/10.1103/PhysRevB.106.235119} {Phys. Rev. B \textbf{106}, 235119 (2022)}.
\bibitem{anupweyl}A. P. Sakhya, C.-Y. Huang, G. Dhakal, X.-J. Gao, S. Regmi, B. Wang, W. Wen, R. H. He, X. Yao, R. Smith et al., \href{https://doi.org/10.1103/PhysRevMaterials.7.L051202} {Phys. Rev. Mater. \textbf{7}, L051202 (2023)}.
\bibitem{YCr6Ge6}Y. Ishii, H. Harima, Y. Okamoto, J.-i. Yamaura, and Z. Hiroi, \href{https://doi.org/10.7566/JPSJ.82.023705} {J. Phys. Soc. Jpn. \textbf{82}, 023705 (2013)}.
\bibitem{ZXShen}G. A. Wigger, F. Baumberger, Z. X. Shen, Z. P. Yin, W. E. Pickett, S. Maquilon, and Z. Fisk, \href{https://doi.org/10.1103/PhysRevB.76.035106} {Phys. Rev. B \textbf{76}, 035106 (2007)}.
\bibitem{Kun2023AV3Sb5} K. Jiang, T. Wu, J.-X. Yin, Z. Wang, M. Z. Hasan, S. D. Wilson, X. Chen, and J. Hu, \href{https://doi.org/10.1093/nsr/nwac199} {Natl. Sci. Rev. \textbf{10}, (2022)}.
\bibitem{Takemi2022KV3Sb5}T. Kato, Y. Li, T. Kawakami, M. Liu, K. Nakayama, Z. Wang, A. Moriya, K. Tanaka, T. Takahashi, Y. Yao et al., \href{https://doi.org/10.1038/s43246-022-00255-1} {Commun. Mater. \textbf{3}, 30 (2022)}.
\bibitem{Soohyun2021RbV3Sb5}S. Cho, H. Ma, W. Xia, Y. Yang, Z. Liu, Z. Huang, Z. Jiang, X. Lu, J. Liu, Z. Liu et al., \href{https://journals.aps.org/prl/abstract/10.1103/PhysRevLett.127.236401} {Phys. Rev. Lett. \textbf{127}, 236401 (2021)}.
\bibitem{Ming}Y. Hu, S. M. L. Teicher, B. R. Ortiz, Y. Luo, S. Peng, L. Huai, J. Ma, N. C. Plumb, S. D. Wilson, J. He et al., \href{https://www.sciencedirect.com/science/article/pii/S2095927321007349}  {Sci. Bull. \textbf{67}, 495 (2022)}.
\bibitem{Chen}L. Chen, Y. Zhou, H. Zhang, X. Ji, K. Liao, Y. Ji, Y. Li, Z. Guo, X. Shen, R. Yu et al., \href{https://doi.org/10.1038/s43246-024-00513-4} {Commun Mater \textbf{5}, 73 (2024)}.
\bibitem{Guo134}J. Guo, L. Zhou, J. Ding, G. Qu, Z. Liu, Y. Du, H. Zhang, J. Li, Y. Zhang, F. Zhou et al.,\href{https://doi.org/10.1016/j.scib.2024.06.036} {Sci. Bull., \textbf{69}, 2660, (2024)}.
\bibitem{Chen134}Z. Zheng, L. Chen, X. Ji, Y. Zhou, G. Qu, M. Hu, Y. Huang, H. Weng, T. Qian, and G. Wang, \href{https://doi.org/10.1007/s11433-023-2344-6} {Sci. China Physics Mech. Astronomy., \textbf{67}, 267411, (2024)}.
\end{thebibliography}
\end{document}